\RequirePackage{booktabs}

\documentclass[sn-mathphys-num]{sn-jnl_modified}

\usepackage{amsmath,amsfonts}
\usepackage{amssymb,amsthm}
\usepackage{makecell}
\usepackage{geometry}

\usepackage{mathtools}%
\usepackage{algorithm}%
\usepackage[noend]{algpseudocode}%
\usepackage{adjustbox}%
\usepackage{xcolor}%
\usepackage{graphicx}%
\usepackage{multirow, multicol}%
\usepackage{mathrsfs}%
\usepackage{booktabs}%
\usepackage{algorithm}%
\usepackage{listings}%
\usepackage{cancel}%
\usepackage{makecell}%
\usepackage{comment}%
\usepackage{hyperref}%
\usepackage{multirow}

\newcommand{\ZZ}{\ensuremath{\mathbb{Z}}}
\newcommand{\NN}{\ensuremath{\mathbb{N}}}
\newcommand{\FF}{\ensuremath{\mathbb{F}}}
\newcommand{\cC}{\ensuremath{\mathcal{C}}}
\newcommand{\cX}{\ensuremath{\mathcal{X}}}

\DeclareFontFamily{U}{mathx}{}
\DeclareFontShape{U}{mathx}{m}{n}{<-> mathx10}{}
\DeclareSymbolFont{mathx}{U}{mathx}{m}{n}
\DeclareMathAccent{\widecheck}{0}{mathx}{"71}

\def\letus{%
	\mathord{\setbox0=\hbox{$\exists$}%
		\hbox{\kern 0.125\wd0%
			\vbox to \ht0{%
				\hrule width 0.75\wd0%
				\vfill%
				\hrule width 0.75\wd0}%
			\vrule height \ht0%
			\kern 0.125\wd0}%
	}%
}

\newtheorem{theorem}{Theorem}[subsection]
\newtheorem{proposition}[theorem]{Proposition}%
\newtheorem{lemma}[theorem]{Lemma}
\newtheorem{corollary}[theorem]{Corollary}

\newtheorem{remark}{Remark} 
\newtheorem{example}{Example}
\newtheorem{definition}[theorem]{Definition}

\newtheorem{mytheorem}{Theorem}
\newtheorem{mylemma}{Lemma}

\begin{document}

\title{Explicit bases for Riemann-Roch spaces on elliptic curves and their application in constructing various elliptic code families}

\author*[1,2]{\fnm{Artyom} \sur{Kuninets}}\email{artkuninets@yandex.ru}

 \author[1,2]{\fnm{Ekaterina} \sur{Malygina}}\email{emalygina@hse.ru}

\affil[1]{\orgdiv{MIEM}, \orgname{HSE University}, \orgaddress{\street{Tallinskaya~Ul.~34}, \city{Moscow}, \postcode{123458}, \country{Russian Federation}}}

\affil[2]{\orgname{QApp}, \orgaddress{\street{Bolshoy~blvd~30b1}, \city{Moscow}, \postcode{121205}, \country{Russian Federation}}}

\abstract{
In this paper, we determine explicit bases for Riemann–Roch spaces associated with various families of elliptic codes. We establish the feasibility and provide exact algorithms for constructing bases of Riemann–Roch spaces corresponding to arbitrary divisors on elliptic curves, including the non-effective case. These results are subsequently applied to derive bases for quasi-cyclic elliptic codes and their subfield subcodes, for the class of Goppa-like elliptic codes, and for quasi-cyclic variants of MDS and isometry-dual (iso-dual) elliptic codes. For algebraic geometry code applications, having an explicit description of Riemann–Roch space bases for arbitrary divisors is particularly valuable as it simultaneously enables efficient code construction and reveals structural properties of the codes, leading, for example, to new cryptanalysis methods when these codes are employed in cryptographic schemes.
}



\maketitle

\section{Author contributions}
All authors contributed to the study conception and design. Material preparation, data collection and analysis were performed by Artyom Andreevich Kuninets and Ekaterina Sergeevna Malygina. The first draft of the manuscript was written by Artyom Andreevich Kuninets and Ekaterina Sergeevna Malygina and all authors commented on previous versions of the manuscript. All authors read and approved the final manuscript.

\section{Introduction}

Riemann–Roch spaces represent fundamental vector spaces of rational functions characterized by prescribed conditions on their zero/pole multiplicities and local behavior. These spaces serve as a foundational tool in contemporary applications of algebraic geometry to diverse computer science disciplines, particularly in coding theory and cryptographic system design. For various applications, it is essential to construct bases of these spaces for divisors of sufficiently large degrees.

The first results on applying geometric methods to derive Riemann–Roch spaces associated with divisors of algebraic curves were obtained by Brill and Noether~\cite{BN74}. The original algorithm was restricted to curves with only ordinary singularities. Le Brigand and Risler extended the method to arbitrary plane curves in~\cite{BR88}. Subsequent developments of geometric approaches appeared in~\cite{HI94,AP20,ABCL22}. In modern computer algebra systems, so-called 'arithmetic' algorithms have become widely implemented. Currently, the best and most widely used algorithm is proposed by Hess~\cite{HESS02}. This algorithm is more universal than geometric methods, handling all types of singularities without requiring generic changes of variables. However, it necessitates computing integral closures of arbitrary orders in the input function field. Since integral closures are generally computationally expensive, this leads to worse complexity compared to geometric approaches.

For cryptographic applications, it is essential to know the explicit form of a basis for the Riemann–Roch space. This knowledge is crucial both for enhancing the efficiency of code-based schemes and for their cryptanalysis. Currently, for most curves, the basis structure has only been determined for divisors of the form $G = kP_\infty$. For the Hermitian curve, initial results characterized the Riemann–Roch space at infinity~\cite{Stich90}, later extended to two rational places~\cite{Park10}. Systematic constructions of bases for these spaces were developed in~\cite{MMP05}. The structure of Riemann–Roch space of Suzuki curves was studied in~\cite{Henn1978}, with coding applications developed for single-point~\cite{HS90} and two-point codes~\cite{Matt04}. Further works included investigations of Ree group curves~\cite{Ped92}, GK maximal curves~\cite{GK09}, and their GGS generalizations~\cite{GDS10}. In~\cite{Ska16} cyclic covers of Suzuki curves are studied. A novel approach for constructing bases of Riemann–Roch spaces for multipoint divisors on generalized Hermitian curves was introduced in~\cite{HZ15}. This methodology was subsequently extended to BBGS curves~\cite{HU17}, Kummer extensions~\cite{HU18}, and GGS curves~\cite{HU17+}.

Elliptic codes and their subcodes exhibit attractive parameters. For instance,~\cite{ZC22} analyzes the parameters of subfield subcodes of one-point elliptic codes, showing that in the binary case their minimum distance surpasses that of comparable-rate BCH codes. However, the basis for the Riemann–Roch space has only been derived for divisors of the form $G = kP_\infty$~\cite{Stich90}, leaving codes associated with multipoint divisors currently unexplored in the case of arbitrary field characteristic.

\textbf{Our contribution.} In this work, we present new results on bases of Riemann–Roch spaces associated with divisors of elliptic curves, enabling the effective construction of such codes. The following Lemma gives explicit basis for one-point divisors.

\begin{mylemma}\label{intro:lem1}
Let $\mathcal{E}/\FF_{p^m}$ be an elliptic curve, and $P = (\alpha, \beta) \in \mathcal{E}(\FF_{p^m})\backslash P_\infty$. The basis of the Riemann – Roch space associated with the divisor $G = kP$, where $k \in \NN_{\geq 2}$, is defined as follows: 
\[
\mathscr{L}_b = \left\{ 1, f_2, f_3, \dots, f_k \right\}, \text{ where } f_s(X,Y) = \frac{Y + A_s(X)}{(X - \alpha)^s}, \quad 2 \leq s \leq k,
\]
and $A_s(X) \in \FF_{p^m}(\mathcal{E})$ is a function of degree $\deg(A_s) \leq s-1$, satisfying:
\begin{enumerate}
    \item $v_{P^\prime}(Y + A_s(X)) \geq s$ for $P^\prime = -P$
    \item $v_Q(Y + A_s(X)) \geq 0$ for any $Q \neq P^\prime$
\end{enumerate}
\end{mylemma}

Using Lemma~\ref{intro:lem1} and Algorithm~\ref{alg:R-R_single_char_2_3}, which describes a specific method for constructing functions, we have obtained a possible basis for an arbitrary multipoint divisor.

\begin{mytheorem}
Let $\mathcal{E}/\mathbb{F}_{p^m}$ be an elliptic curve and let $P_1 = (\alpha_1, \beta_1), \dots, P_z = (\alpha_z, \beta_z)$ be distinct rational points in $\mathcal{E}(\mathbb{F}_{p^m}) \setminus \{P_\infty\}$. For a divisor $G = \sum_{i=1}^z k_i P_i$ with $k_i \in \mathbb{N}_{\geq 2}$, a basis for the Riemann–Roch space $\mathscr{L}(G)$ is given by:
\[
\mathscr{L}_b = \{1\} \cup \left\{ f_{i,s} \mid 1 \leq i \leq z,\ 2 \leq s \leq k_i \right\} \cup \left\{ g_i \mid 1 \leq i \leq z-1 \right\},
\]
where the functions are defined as follows:
\[
f_{i,s}(X,Y) = \frac{Y + A_{i,s}(X)}{(X - \alpha_i)^s}, \quad 2 \leq s \leq k_i,
\]
and
\[
g_i(X,Y) = 
\begin{cases}
\dfrac{Y + B_i(X)}{(X - \alpha_i)(X - \alpha_{i+1})}, & \text{if } \alpha_i \neq \alpha_{i+1},\\[8pt]
\dfrac{1}{X - \alpha_i}, & \text{if } \alpha_i = \alpha_{i+1}.
\end{cases}
\]
Here, $A_{i,s}(X) \in \mathbb{F}_{p^m}[X]$ has degree $\deg(A_{i,s}) \leq s-1$ and $B_i(X) = \left( \frac{\beta_{i+1} - \beta_i}{\alpha_{i+1} - \alpha_i} + a_1 \right) X + \left( (\beta_i + a_1\alpha_i + a_3) - \left( \frac{\beta_{i+1} - \beta_i}{\alpha_{i+1} - \alpha_i} + a_1 \right) \alpha_i \right)$.

\end{mytheorem}

Finally, building upon our previous results, we derive explicit bases for Riemann–Roch spaces associated with two particularly cryptographically significant code families: quasi-cyclic \cite{TW67} elliptic codes and Goppa-like \cite{KLN23} codes. We also consider new constructions of MDS and isometry-dual (iso-dual) elliptic codes \cite{ZZ25}, namely their quasi-cyclic variants.

\textbf{Organization of the paper.} Section~\ref{sec:background} introduces preliminary information necessary for understanding the construction of Riemann–Roch spaces and algebraic geometry codes. Next, in Section~\ref{sec:R-R_spaces} we present detailed algorithm for constructing various families of elliptic codes with arbitrary divisors. Sections~\ref{sec:QC} and~\ref{sec:Goppa-like} focus specifically on the construction of quasi-cyclic elliptic codes and Goppa-like elliptic codes, respectively, with particular emphasis on their potential cryptographic applications. In Section \ref{sec:mds_iso-dual} we consider the construction of quasi-cyclic variants of MDS and iso-dual elliptic codes associated with divisors $G \neq kP_\infty$.

\section{Preliminaries}\label{sec:background}

This section does not contain new results, it consists of relevant known facts from coding theory necessary for proofs and understanding of our work.

\subsection{Elliptic curves and their automorphisms}

The elliptic curve $\mathcal{E}$ over the field $\FF_q$ is defined in affine coordinates by the smooth equation  
\begin{equation*}
    y^{2} + a_{1} xy + a_{3} y = x^{3} + a_{2} x^{2} + a_{4} x + a_{6},
\end{equation*}
where $a_i \in \FF_q$. This equation is called the \textit{full Weierstrass form} in affine coordinates.

In this case the genus of the curve $g = 1$ and the set of \textit{rational points of the elliptic curve} $\mathcal{E}(\FF_q)$ is defined as  
\[
\mathcal{E}(\FF_q) = \left\{(\alpha, \beta) \in \mathbb{F}_{q}^{2}: \beta^{2} + a_{1} \alpha \beta + a_{3} \beta = \alpha^{3} + a_{2} \alpha^{2} + a_{4} \alpha + a_{6}\right\} \cup \{P_{\infty}\}.
\]

The number of rational points on the elliptic curve satisfies the \textit{Hasse bound}  
\begin{equation}\label{EC:optimal}
    q + 1 - \lfloor 2\sqrt{q} \rfloor \leq \left|\mathcal{E}(\FF_q)\right| \leq q + 1 + \lfloor 2\sqrt{q} \rfloor.
\end{equation}
If the number of rational points on curve achieves higher or lower bound, then the curve is called \textit{optimal}. Furthermore, the curve $\mathcal{E}$ is called \textit{maximal} (resp. \textit{minimal}) if it attains the upper (resp. lower) bound in~\eqref{EC:optimal}.

Next, we present known results on the structure of automorphism groups of elliptic curves. Let $\operatorname{Aut}(\mathcal{E}/\overline{\mathbb{F}}_q, P_\infty)$ denote the automorphism group of elliptic curve fixing infinity point, i.e. 
\[
\operatorname{Aut}(\mathcal{E}/\overline{\mathbb{F}}_q, P_\infty) = \{ \sigma \in \operatorname{Aut}(\mathcal{E}/\overline{\mathbb{F}}_q) : \sigma(P_\infty) = P_\infty\}.
\]
\begin{theorem}[{\cite[Theorem 10.1]{Silverman09}}]
    \label{Aut_EC:Aut_theorem2}
    Let $\mathcal{E}/\mathbb{F}_q$ be an elliptic curve, and let $j(\mathcal{E}) \in \mathbb{F}_q$ denote its $j$-invariant. Then the order of the automorphism group $\operatorname{Aut}(\mathcal{E}/\overline{\mathbb{F}}_q, P_\infty)$ over the algebraic closure is:
    \begin{enumerate}
        \item $|\operatorname{Aut}(\mathcal{E}/\overline{\mathbb{F}}_q, P_\infty)| = 2$ if $j(\mathcal{E}) \neq 0, 1728$;
        \item $|\operatorname{Aut}(\mathcal{E}/\overline{\mathbb{F}}_q, P_\infty)| = 4$ if $j(\mathcal{E}) = 1728$ and $\operatorname{char}(\mathbb{F}_q) \notin \{2, 3\}$;
        \item $|\operatorname{Aut}(\mathcal{E}/\overline{\mathbb{F}}_q, P_\infty)| = 6$ if $j(\mathcal{E}) = 0$ and $\operatorname{char}(\mathbb{F}_q) \notin \{2, 3\}$;
        \item $|\operatorname{Aut}(\mathcal{E}/\overline{\mathbb{F}}_q, P_\infty)| = 12$ if $j(\mathcal{E}) \in \{0, 1728\}$ and $\operatorname{char}(\mathbb{F}_q) = 3$;
        \item $|\operatorname{Aut}(\mathcal{E}/\overline{\mathbb{F}}_q, P_\infty)| = 24$ if $j(\mathcal{E}) \in \{0, 1728\}$ and $\operatorname{char}(\mathbb{F}_q) = 2$.
    \end{enumerate}   
\end{theorem}
The following classification gives the structure of full automorphism groups of elliptic curves (not restricted to automorphisms fixing the point at infinity).

\begin{theorem}[{\cite{Silverman09}}]\label{Aut_EC:Aut_remark}
    The automorphism group of the elliptic curve $\operatorname{Aut}(\mathcal{E}/\overline{\mathbb{F}}_q)$ is isomorphic to:
    \[
        \begin{array}{rll}
            T_E \rtimes (\mathbb{Z}/2\mathbb{Z}), & \text{if} & j \neq 0, 1728; \\
            T_E \rtimes (\mathbb{Z}/4\mathbb{Z}), & \text{if} & j = 1728 \text{ and } \operatorname{char}(\mathbb{F}_q) \notin \{2, 3\}; \\
            T_E \rtimes (\mathbb{Z}/6\mathbb{Z}), & \text{if} & j = 0 \text{ and } \operatorname{char}(\mathbb{F}_q) \notin \{2, 3\}; \\
            T_E \rtimes (\mathbb{Z}/4\mathbb{Z} \rtimes \mathbb{Z}/3\mathbb{Z}), & \text{if} & j = 0 = 1728 \text{ and } \operatorname{char}(\mathbb{F}_q) = 3; \\
            T_E \rtimes \left(Q_8 \rtimes \mathbb{Z}/3\mathbb{Z}\right), & \text{if} & j = 0 = 1728 \text{ and } \operatorname{char}(\mathbb{F}_q) = 2,
        \end{array}
    \]
where $T_E$ denotes the translation group of elliptic function field $E \simeq \FF_q(\mathcal{E})$.
\end{theorem}

\subsection{Algebraic geometry codes}

Let $\mathcal{X}$ be a smooth, projective, and absolutely irreducible algebraic curve of genus $g$ defined over the finite field $\mathbb{F}_{q}$. A \emph{divisor} on $\mathcal{X}$ (defined over $\mathbb{F}_{q}$) is a finite formal sum of places of $\mathcal{X}$, that is,
\[
G = \sum_{P} v_P(G)P,
\]
where $v_P(G) \in \mathbb{Z}$ and all but finitely many coefficients are zero. The set of all such divisors forms an abelian group, denoted by $\operatorname{Div}(\mathcal{X})$. We denote by $\mathcal{X}(\mathbb{F}_{q})$ the set of $\mathbb{F}_{q}$-rational points of $\mathcal{X}$.

For a divisor $G = \sum_P v_P(G)P$, its \emph{support} is defined as
\[
\operatorname{Supp}(G) = \{P \mid v_P(G) \neq 0\},
\]
and its \emph{degree} is given by
\[
\deg(G) = \sum_P v_P(G)\deg(P).
\]

A divisor $G$ is said to be \emph{effective}, written $G \geq 0$, if $v_P(G) \geq 0$ for every place $P$. This induces a partial order on $\operatorname{Div}(\mathcal{X})$ by declaring $G_1 \geq G_2$ whenever $G_1 - G_2$ is effective, that is, whenever $v_P(G_1) \geq v_P(G_2)$ for all $P$.

The \textit{function field} of the curve $\mathcal{X}$ is denoted by
\[
\mathbb{F}_{q}(\mathcal{X})=\left\{ \frac{g(x,y)}{h(x,y)} \mid g,h \in \mathbb{F}_{q}[x, y], h \neq 0 \right\}/\sim,
\]
where $\frac{g}{h} \sim \frac{g^\prime}{h^\prime} \Leftrightarrow (gh^\prime - g^\prime h)(P) = 0 \,\forall P \in \mathcal{X}$.

For a nonzero rational function $f \in \mathbb{F}_{q}(\mathcal{X})$, its \emph{principal divisor} is defined by
\[
(f) = (f)_0 - (f)_\infty,
\]
where $(f)_0$ and $(f)_\infty$ denote the zero and pole divisors of $f$, respectively.

Given a divisor $G \in \operatorname{Div}(\mathcal{X})$, the associated \emph{Riemann–Roch space} is defined as
\[
\mathscr{L}(G)
=
\{f \in \mathbb{F}_{q}(\mathcal{X}) \mid (f) + G \geq 0 \}
\cup \{0\}.
\]
This is a finite-dimensional vector space over $\mathbb{F}_{q}$, and its dimension is denoted by $\ell(G)$.

Let $D = P_1 + \cdots + P_n$ be a divisor consisting of $n$ distinct $\mathbb{F}_{q}$-rational points of $\mathcal{X}$, and assume that $\operatorname{Supp}(D) \cap \operatorname{Supp}(G) = \varnothing$.

We define the evaluation map
\[
\operatorname{ev}_D :
\begin{cases}
\mathscr{L}(G) \longrightarrow \mathbb{F}_{q}^n, \\
f \longmapsto \bigl(f(P_1), \dots, f(P_n)\bigr).
\end{cases}
\]

\begin{definition}[Algebraic Geometry Code]
The \emph{algebraic geometry code} associated with the pair $(D,G)$ is defined as
\[
\mathcal{C}_{\mathscr{L}}(D,G)
=
\{ \operatorname{ev}_D(f) \mid f \in \mathscr{L}(G) \}
\subseteq \mathbb{F}_{q}^n.
\]
\end{definition}

The code $\mathcal{C}_{\mathscr{L}}(D,G)$ is a linear code with parameters $[n,k,d]$, where:
\begin{itemize}
    \item $n = \deg(D)$ is the length of the code,
    \item $k = \dim_{\mathbb{F}_{q}} \mathscr{L}(G)$ is its dimension,
    \item $d$ is its minimum Hamming distance.
\end{itemize}

By the Riemann–Roch theorem (see, e.g., \cite[Theorem 2.2.2]{Stichtenoth09}), the parameters satisfy:
\[
k \geq \deg(G) + 1 - g,
\qquad
d \geq n - \deg(G).
\]
Moreover, if $2g - 2 < \deg(G) < n$, then $k = \deg(G) + 1 - g$.

If $\{f_1, \ldots, f_k\}$ is a basis of $\mathscr{L}(G)$, then a generator matrix of $\mathcal{C}_{\mathscr{L}}(D,G)$ is given by
\[
\begin{pmatrix}
f_1(P_1) & f_1(P_2) & \cdots & f_1(P_n) \\
f_2(P_1) & f_2(P_2) & \cdots & f_2(P_n) \\
\vdots & \vdots & \ddots & \vdots \\
f_k(P_1) & f_k(P_2) & \cdots & f_k(P_n)
\end{pmatrix}.
\]

In addition to the evaluation code $\mathcal{C}_{\mathscr{L}}(D,G)$, one can associate to the divisors $D$ and $G$ a second code constructed from differentials, denoted by $\mathcal{C}_{\Omega}(D,G)$. Let $\Omega$ denote the space of rational (Weil) differentials on the curve $\mathcal{X}$. This space is one-dimensional over $\mathbb{F}_{q}$. For each nonzero differential $\omega \in \Omega$, we denote by $(\omega)$ its corresponding divisor.

For a divisor $G \in \operatorname{Div}(\mathcal{X})$, define the space
\[
\Omega(G) = \{\omega \in \Omega \mid \omega = 0 \ \text{or} \ (\omega) \geq G\}.
\]
This is the set of differentials whose divisor is bounded below by $G$.

The \emph{differential algebraic geometry code} associated with $(D,G)$ is then defined by
\[
\mathcal{C}_{\Omega}(D,G)
=
\left\{
\bigl(\operatorname{Res}_{P_1}(\omega), \ldots, \operatorname{Res}_{P_n}(\omega)\bigr)
\;\middle|\;
\omega \in \Omega(G-D)
\right\},
\]
where $\operatorname{Res}_{P_i}(\omega)$ denotes the residue of $\omega$ at the place $P_i$.

\medskip

The connection between evaluation codes and differential codes is described in the following result.

\begin{proposition}[{\cite[Proposition 2.2.10]{Stichtenoth09}}]
Let $\mathcal{C}_{\mathscr{L}}(D,G)$ be the algebraic geometry code defined on the curve $\mathcal{X}$. Then its dual code satisfies
\[
\mathcal{C}_{\mathscr{L}}(D,G)^\perp
=
\mathcal{C}_{\mathscr{L}}(D,G^\perp)
=
\mathcal{C}_{\Omega}(D,G),
\]
where
\[
G^\perp = D - G + (\eta),
\]
and $\eta$ is a Weil differential such that
\[
v_P(\eta) = -1
\quad \text{and} \quad
\operatorname{Res}_{P}(\eta) = 1
\]
for every $P \in \operatorname{Supp}(D)$.
\end{proposition}

For more details on function fields and algebraic geometry codes we refer to the book~\cite{Stichtenoth09}.

\section{On constructing elliptic codes with arbitrary divisors} \label{sec:R-R_spaces}

In this section, we present bases for Riemann–Roch spaces used in constructing families of elliptic codes from~\cite{Bar,KLN23}. Note that in most existing works, authors primarily consider codes associated with divisors of the form $G = kP_\infty$. The basis of the Riemann–Roch space in this case is trivial and consists of functions of the form:
\[
\mathscr{L}_b(kP_\infty) = \{x^iy^j \mid 2i+3j \leq k,j = \overline{0, 1}\}.
\]

In this work, we provide specific algorithm for constructing bases of Riemann–Roch spaces associated with arbitrary divisors $G = \sum_{i=1}^z k_i P_i$ with $k_i \in \mathbb{N}_{>0}$ on elliptic curves. This result can be used both for efficient construction of codes associated with such divisors and for investigating cryptographic properties of elliptic codes and their subcodes with the aim of embedding them into code-based cryptosystems.

\subsection{Bases of Riemann–Roch spaces associated with divisors $G \neq kP_\infty$ in general case}\label{sec:R-R-General_case}

\subsubsection{Bases for effective divisors}

\begin{lemma}\label{lem:R-R_single}
Let $\mathcal{E}/\FF_{p^m}$ be an elliptic curve, and let $P = (\alpha, \beta) \in \mathcal{E}(\FF_{p^m})\backslash P_\infty$. The basis $\mathscr{L}_b$ of the Riemann – Roch space associated with the divisor $G = kP$, where $k \in \NN_{\geq 2}$, is defined as follows: 

Case 1: If $P \not\in \mathcal{E}[2]$ (i.e., $-P \neq P$), then the basis is:
\[
\mathscr{L}_b = \left\{ 1, f_2, f_3, \dots, f_k \right\}, \qquad 
f_s(X,Y) = \frac{Y + A_s(X)}{(X - \alpha)^s},
\]
where $A_s(X) \in \FF_{p^m}(\mathcal{E})$ is a function of degree $\deg(A_s) \leq s-1$, satisfying:
\begin{enumerate}
    \item $v_{P^\prime}(Y + A_s(X)) \geq s$ for $P^\prime = -P$;
    \item $v_Q(Y + A_s(X)) \geq 0$ for any $Q \neq P^\prime$.
\end{enumerate}

Case 2: If $P \in \mathcal{E}[2]$ (i.e., $-P = P$), then the basis is:
\[
\mathscr{L}_b = \left\{ 1 \right\} \cup \left\{ \frac{1}{(X-\alpha)^s} \;\middle|\; 1 \leq s \leq \left\lfloor \frac{k}{2} \right\rfloor \right\} \cup \left\{ \frac{Y-\beta}{(X-\alpha)^s} \;\middle|\; 2 \leq s \leq \left\lfloor \frac{k+1}{2} \right\rfloor \right\}.
\]

\begin{proof}

In the case of elliptic function fields $\FF_q(\mathcal{E})$, the local uniformizing parameter at the point $P_{\alpha, \beta}$ is the function $t = X - \alpha$. For the function $f = \frac{1}{X - \alpha}$ and any $s \geq 2$, the following holds:
\[
\operatorname{div}(f^s) = (f^s)_0 - (f^s)_\infty = 2s P_\infty - sP_{\alpha, \beta} - sP_{\alpha, \beta^\prime}, \text{ since } v_{P_\infty}(X^iY^j) = -2i - 3j.
\]

However, to construct a basis for the Riemann–Roch space, the function must have a pole only at the point $P$. Therefore, it is necessary to construct a function for which the point $ P^\prime = P_{\alpha, \beta^\prime} $ is not a pole simultaneously with $ P $. In the case $P \in \mathcal{E}[2]$, the function has a pole of order 2 at point $P$, since $-P = P$. Therefore, we will only consider points that are not 2-torsion points.

Taking into account that $v_P(XY)=v_P(X)+v_P(Y)$, to obtain function $f_s(X,Y)$ without pole at $P^\prime$, and which satisfies the condition $\operatorname{div}(f_s) \geq - G$, we can multiply $f^s = \frac{1}{(X-\alpha)^s}$ by any function $g(X, Y)$, such that
\[
\deg(g) \leq s, \quad v_{{P}^\prime}(g) \geq s \text{ for any fixed } s \geq 2.
\]

Consider functions of the form $g(X,Y) = Y + A_s(X)$, where $A_s(x) = \sum_{j = 0}^{s-1}c_jX^{j}$. Then the following holds:
\[
\operatorname{div}\left(\frac{Y + A_s(X)}{(X - \alpha)^s}\right) = 2sP_\infty - sP_{\alpha, \beta} - sP_{\alpha, \beta^\prime} + \operatorname{div}\left(Y+A_s(X)\right)
\]
\[
\begin{split}
   \text{div}\left(Y + A_s(X)\right) &= (Y + A_s(X))_0 - (Y + A_s(X))_\infty\\ 
   & = (Y + A_s(X))_0 - \max\{v_{P_\infty}(Y), v_{P_\infty}(A_s(X))\}\cdot P_\infty \\
   & \geq (Y + A_s(X))_0 - 2sP_\infty \\
   & \geq sP_{\alpha, \beta^\prime} - 2(s-1)P_\infty + \ldots 
\end{split}
\]
Thus, if $ v_{P'}(Y + A_s(X)) \geq s $, we get: 
\[
\operatorname{div}\left(\frac{Y + A_s(X)}{(X - \alpha)^s}\right) \geq 2sP_\infty - sP_{\alpha, \beta} - sP_{\alpha, \beta^\prime} + sP_{\alpha, \beta^\prime} - 2(s-1)P_\infty \geq -sP_{\alpha, \beta}.
\]

In this case, the zero divisor of the function may include other points, however, for computing the basis of the Riemann–Roch space of a one-point divisor, this fact is not crucial, since the constraints apply only to the poles of functions in $\mathscr{L}(G)$.

Clearly, functions of the form $f_s(X,Y) = \frac{Y + A_s(X)}{(X - \alpha)^s}$ are linearly independent, and therefore they form the basis of $\mathscr{L}(G)$.

For a 2‑torsion point we have $-P = P$. The functions $\frac{1}{(X-\alpha)^s}$ and $\frac{Y-\beta}{(X-\alpha)^s}$ satisfy
\[
\operatorname{div}\!\left(\frac{1}{(X-\alpha)^s}\right) = 2s P_\infty - 2s P,
\qquad
\operatorname{div}\!\left(\frac{Y-\beta}{(X-\alpha)^s}\right) = (2s-3)P_\infty - (2s-1)P + \sum_{Q\in\mathcal{E}[2]\setminus\{P\}} Q.
\]
For $s \geq 2$ the second function has a pole of order $2s-1$ at $P$ and no other poles. These functions provide all pole orders up to $k$, and their number is $1 + \lfloor k/2\rfloor + (\lfloor (k+1)/2\rfloor - 1) = k$, matching $\dim(\mathscr{L}(kP))$.

\end{proof}

\end{lemma}

In Algorithm~\ref{alg:R-R_single_char_2_3}, we consider specific method for constructing functions of the form $Y + A_s(X)$ from Lemma~\ref{lem:R-R_single}. Let us briefly describe the  idea of the algorithm.

Consider the function $h(X,Y) = Y + A_s(X)$. For the function $h(X)$ to have a zero of multiplicity $s+1$ at the point $P_{\alpha, \beta^\prime}$, it is sufficient that the following conditions hold:
\[
\left.\frac{d^{i}}{dX^{i}} \left[ A_{s}(X) + Y \right]\right|_{X=\alpha, Y=\beta^\prime} = 0, \quad i = 0, \ldots, s-1,
\]
i.e., the first $s-1$ derivatives of $Y + A_s(X)$ must vanish at $P^\prime$.

This can be achieved by expanding the elliptic curve into a Taylor series around the point $P^\prime$ using the local parameter $t = X - \alpha$. Thus, we obtain a function of the form:
\[
Y = \beta^\prime + \sum_{j = 1}^{s-1}c_jt^j,
\]
where $c_j$ are the coefficients in the Taylor expansion of the elliptic curve. The sum $\sum_{j = 1}^{s-1}c_jt^j$ ensures that the first $s-1$ derivatives vanish at $P^\prime$.


Characteristic $2$: For elliptic curves defined over fields of characteristic 2, we begin with the general Weierstrass equation:
\[
Y^2 + a_1XY + a_3Y = X^3 + a_2X^2 + a_4X + a_6.
\]
To derive the coefficient $c_1$ in the Taylor expansion around the conjugate point $P^\prime = (\alpha, \beta^\prime)$, we perform implicit differentiation of the curve equation with respect to $X$. This yields:
\[
2Y \frac{dY}{dX} + a_1Y + a_1X \frac{dY}{dX} + a_3 \frac{dY}{dX} = 3X^2 + 2a_2X + a_4,  
\]
\[
a_1Y + a_1X \frac{dY}{dX} + a_3 \frac{dY}{dX} = X^2 + a_4.
\]

Considering the Taylor expansion $Y = \beta^\prime + c_1t + \cdots$, where $t = X - \alpha$, we evaluate the derivative at $t=0$ (i.e., at point $P^\prime$) to obtain $\frac{dY}{dX}\big|_{t=0} = c_1$. Substituting $X = \alpha$ and $Y = \beta^\prime$ into the differentiated equation and solving for $c_1$ gives:
\[
c_1 = \frac{\alpha^2 + a_4 + a_1\beta^\prime}{a_1\alpha + a_3}, \,\text{ where }\,  a_1\alpha + a_3 \neq 0\, \text{ if } P \notin \mathcal{E}[2].
\]

Characteristic $> 3$: When working in characteristic greater than 3, the elliptic curve equation simplifies to the short Weierstrass form:
\[
Y^2 = X^3 + a_4X + a_6.
\]
Following a similar approach, we differentiate implicitly to obtain:
$
2Y \frac{dY}{dX} = 3X^2 + a_4.
$
Again using the Taylor expansion at the conjugate point $P^\prime = (\alpha, \beta^\prime)$ and evaluating at $t=0$, we solve for $c_1$ to find:
\[
c_1 = \frac{3\alpha^2 + a_4}{2\beta^\prime}.
\]

\begin{algorithm}[ht!]
\caption{Basis for $\mathscr{L}(kP_{\alpha,\beta})$}
\label{alg:R-R_single_char_2_3}
\begin{algorithmic}[1]
\Require $\mathcal{E}/\FF_{q}:\left\{\begin{array}{cl}
Y^2 + a_1XY + a_3Y = X^3 + a_2X^2 + a_4X + a_6,      &  \text{if } \operatorname{char}(\FF_q) = 2 \\
 Y^2 = X^3 + a_4X + a_6,    &  \text{if } \operatorname{char}(\FF_q) \geq 3 
\end{array}\right.$  

Point $P_{\alpha, \beta} = (\alpha, \beta) \in \mathcal{E}(\mathbb{F}_{p^m})\backslash \mathcal{E}[2]$,

Integer $k \geq 2$
\Ensure Basis $\{1, f_2, \dots, f_k\}$ for $\mathscr{L}(kP_{\alpha, \beta})$

    \State $\beta^\prime \gets \begin{cases} 
        \beta + a_1\alpha + a_3  & \operatorname{char}(\FF_q) = 2\\
        -\beta & \operatorname{char}(\FF_q) \geq 3
    \end{cases}$
\State $\mathscr{L}_b \gets \{1\}$ 
\State Set local parameter $t \gets X - \alpha$

\For{$s = 2$ \textbf{to} $k$}\Comment{ Expand $Y = \beta^\prime + \sum_{j=1}^{s-1} c_j t^j + \mathcal{O}(t^s)$}
\State $c_1 \gets \left\{
\begin{array}{cl}
        \dfrac{\alpha^2 + a_4 + a_1\beta^\prime}{a_1\alpha + a_3} & \operatorname{char}(\FF_q) = 2
        \\[2.5ex]
        \dfrac{3\alpha^2 + a_4}{2\beta^\prime} & \operatorname{char}(\FF_q) \geq 3
    \end{array}\right.
    $
    \For{$j = 2$ \textbf{to} $s$} \Comment{Substitute $X = t + \alpha$ into the curve equation}
    
        \State $S_j \gets \sum\limits_{i=1}^{j-1} c_i c_{j-i}$
        \State $c_j = \left\{
\begin{array}{cl}
            \dfrac{(\alpha + a_2)\delta_{j,2} + \delta_{j,3} + a_1c_{j-1} + S_j}{a_1\alpha + a_3} & \operatorname{char}(\FF_q) = 2 \\[2ex]
            \dfrac{(3\alpha)\delta_{j,2} + \delta_{j,3} - S_j}{2\beta'} & \operatorname{char}(\FF_q) \geq 3
\end{array}\right. \quad \text{\Comment{$\delta_{i,j}$ –- Kronecker delta function}}$
    \EndFor 
    \State $A_s(X) \gets -\beta^\prime - \sum\limits_{j=1}^{s-1} c_j (X - \alpha)^j$
    \State $f_s \gets \dfrac{Y + A_s(X)}{(X - \alpha)^s}$
    \State $\mathscr{L}_b \gets \mathscr{L}_b \cup \{f_s\}$
\EndFor
\State \Return $\mathscr{L}_b$
\end{algorithmic}
\end{algorithm}

\begin{remark}\label{rem:complexity_RR}
Algorithm~\ref{alg:R-R_single_char_2_3} relies on the recursive computation of the coefficients $c_j$ in the local expansion $Y = \beta' + \sum_{j \geq 1} c_j t^j.$ For a fixed $s$, computing $c_1,\dots,c_s$ requires, for each $j$, evaluation of the convolution sum $S_j = \sum_{i=1}^{j-1} c_i c_{j-i}$, which takes $\mathcal{O}(j)$ field operations. Therefore, the total cost of computing all coefficients up to order $s$ is $\sum_{j=1}^{s} \mathcal{O}(j) = \mathcal{O}(s^2)$. Summing over all $s = 2,\dots,k$, we obtain the overall complexity $\sum_{s=2}^{k} \mathcal{O}(s^2) = \mathcal{O}(k^3)$ field operations.

However, the algorithm admits a natural optimization. Since the coefficients $c_j$ do not depend on $s$, they can be computed once up to order $k$ and reused in all subsequent steps. With this optimization, the total cost of computing all coefficients becomes $\mathcal{O}(k^2)$. Moreover, if we store the previous values, moving from $s$ to $s+1$ only requires adding the term $c_{s}(X-\alpha)^{s}$ to the polynomial $A_s(X)$. Consequently, after precomputing the $c_j$, each $f_s$ can be constructed in $\mathcal{O}(1)$ time. Hence the overall complexity of Algorithm~\ref{alg:R-R_single_char_2_3} is $\mathcal{O}(k^2)$, dominated by the computation of the coefficients $c_j$.
\end{remark}

\begin{theorem}\label{th:R-R_Arbitrary}
Let $\mathcal{E}/\mathbb{F}_{p^m}$ be an elliptic curve in general Weierstrass form:
\[
\mathcal{E}: Y^2 + a_1XY + a_3Y = X^3 + a_2X^2 + a_4X + a_6,
\]
and let $P_1 = (\alpha_1, \beta_1), \dots, P_z = (\alpha_z, \beta_z)$ be distinct rational points in $\mathcal{E}(\mathbb{F}_{p^m}) \setminus \{P_\infty\}$. For a divisor $G = \sum_{i=1}^z k_i P_i$ with $k_i \in \mathbb{N}_{\geq 2}$, a basis for the Riemann–Roch space $\mathscr{L}(G)$ is given by:
\[
\mathscr{L}_b = \{1\} \cup \left\{ f_{i,s} \mid 1 \leq i \leq z,\ 2 \leq s \leq k_i \right\} \cup \left\{ g_i \mid 1 \leq i \leq z-1 \right\},
\]
where the functions are defined as follows.

\noindent 1. Single-point basis functions for $\mathscr{L}(k_iP_i)$ (poles at $P_i$):
\begin{itemize}
\item If $P_i \notin \mathcal{E}[2]$ (i.e., $-P_i \neq P_i$), then for $2 \leq s \leq k_i$,
\[
f_{i,s}(X,Y) = \frac{Y + A_{i,s}(X)}{(X - \alpha_i)^s},
\]
where $A_{i,s}(X) \in \mathbb{F}_{p^m}[X]$ has degree $\leq s-1$ and satisfies $v_{-P_i}(Y + A_{i,s}(X)) \geq s$ (constructed as in Algorithm~\ref{alg:R-R_single_char_2_3}).
\item If $P_i \in \mathcal{E}[2]$ (i.e., $-P_i = P_i$), then for $2 \leq s \leq k_i$,
\[
f_{i,s}(X,Y) = \begin{cases}
\dfrac{1}{(X-\alpha_i)^{s/2}}, & \text{if } s \text{ is even},\\[8pt]
\dfrac{Y - \beta_i}{(X-\alpha_i)^{(s+1)/2}}, & \text{if } s \text{ is odd}.
\end{cases}
\]
(For odd $s=3,5,\dots$, the exponent $(s+1)/2 \geq 2$ guarantees regularity at infinity.)
\end{itemize}

\noindent 2. {Double-point basis functions} for $ \mathscr{L}(k_iP_i + k_{i+1}P_{i+1})$ (simple poles at $P_i$ and $P_{i+1}$): For $1 \leq i \leq z-1$,
\[
g_i(X,Y) = \begin{cases}
\dfrac{Y + B_i(X)}{(X - \alpha_i)(X - \alpha_{i+1})}, & \text{if } \alpha_i \neq \alpha_{i+1},\\[8pt]
\dfrac{1}{X - \alpha_i}, & \text{if } \alpha_i = \alpha_{i+1},
\end{cases}
\]
where $B_i(X)$ is the linear polynomial
\[
B_i(X) = \left( \frac{\beta_{i+1} - \beta_i}{\alpha_{i+1} - \alpha_i} + a_1 \right) X + \left( (\beta_i + a_1\alpha_i + a_3) - \left( \frac{\beta_{i+1} - \beta_i}{\alpha_{i+1} - \alpha_i} + a_1 \right) \alpha_i \right).
\]

\begin{proof}
 
We establish that $\mathscr{L}_b$ forms a basis for $\mathscr{L}(G)$ by demonstrating that every function in $\mathscr{L}_b$ satisfies $\operatorname{div}(f) + G \geq 0$ and all functions are linear independent. Throughout the proof, we denote the conjugate of a point $P_i = (\alpha_i, \beta_i)$ as $-P_i = (\alpha_i, -\beta_i - a_1\alpha_i - a_3)$, consistent with the group law on $\mathcal{E}$.

For the constant function $1$, the divisor is $\operatorname{div}(1) = 0$, so $\operatorname{div}(1) + G = G \geq 0$ since all $k_i > 0$. For each single-point basis function $f_{i,s}(X,Y) = \frac{Y + A_{i,s}(X)}{(X - \alpha_i)^s}$ ($2 \leq s \leq k_i$), Lemma~\ref{lem:R-R_single} guarantees:
\begin{equation*}
v_{P_i}(f_{i,s}) = -s \geq -k_i,
\quad v_{-P_i}(f_{i,s}) \geq 0, 
\quad v_Q(f_{i,s}) \geq 0 \quad \text{for all } Q \notin \{P_i, -P_i\}.
\end{equation*}
Since $G = \sum_{i=1}^z k_i P_i$, we have $\operatorname{div}(f_{i,s}) + G \geq 0$.

For the double-point functions $g_i$, we consider two cases. When $\alpha_i \neq \alpha_{i+1}$, we have:
\[
g_i(X,Y) = \frac{Y + B_i(X)}{(X - \alpha_i)(X - \alpha_{i+1})},
\]
where $B_i(X)$ is constructed to vanish at $-P_i$ and $-P_{i+1}$. The line $B_i(X)$, passing through the points $-P_i$ and $-P_{i+1}$, is given by the equation:
\[
\left( \frac{\beta_{i+1} - \beta_i}{\alpha_{i+1} - \alpha_i} + a_1 \right) X + \left( (\beta_i + a_1\alpha_i + a_3) - \left( \frac{\beta_{i+1} - \beta_i}{\alpha_{i+1} - \alpha_i} + a_1 \right) \alpha_i \right).
\]
This function obviously has zeros at the points $-P_i$ and $-P_{i+1}$, as well as at a third point $Q \not\in \{-P_i, -P_{i+1}\}$, which does not affect the correctness of the basis construction. This implies:
\begin{equation*}
v_{P_i}(g_i) = -1 \geq -k_i, \;
v_{P_{i+1}}(g_i) = -1 \geq -k_{i+1}, \;
v_{-P_i}(g_i) \geq 0, \;
v_{-P_{i+1}}(g_i) \geq 0.
\end{equation*}
At other points $Q \neq P_i, P_{i+1}$, the denominator doesn't vanish unless $x = \alpha_i$ or $x = \alpha_{i+1}$, but at these $x$-values, the numerator vanishes precisely when $y = -\beta_i - a_1\alpha_i - a_3$ or $y = -\beta_{i+1} - a_1\alpha_{i+1} - a_3$ (i.e., at $-P_i$ or $-P_{i+1}$), so $v_Q(g_i) \geq 0$ elsewhere. When $\alpha_i = \alpha_{i+1}$, we use $g_i(X,Y) = \frac{1}{X - \alpha_i}$, which has simple poles at both $P_i$ and $P_{i+1}$ (since they share the same $x$-coordinate but different $y$-coordinates), and is regular elsewhere.

Now we will show linear independence of such functions. Consider linear dependence relation:
\begin{equation} \label{eq:lin-dep}
c_0 + \sum_{i=1}^z \sum_{s=2}^{k_i} c_{i,s} f_{i,s} + \sum_{i=1}^{z-1} d_i g_i = 0.
\end{equation}

The highest possible pole order at $P_j$ is $k_j$, achieved only by $f_{j,k_j}$. Evaluating the Laurent series at $P_j$ and extracting the coefficient of $(x - \alpha_j)^{-k_j}$ gives $c_{j,k_j} = 0$. We proceed inductively: after setting $c_{j,\ell} = 0$ for $\ell > s$, the coefficient of $(x - \alpha_j)^{-s}$ implies $c_{j,s} = 0$. Continuing this descent from $s = k_j$ to $s = 2$, we eliminate all $f_{i,s}$ coefficients.

After removing the $f_{i,s}$ terms, \eqref{eq:lin-dep} simplifies to:
\begin{equation}\label{eq:linear_ind_proof}
 c_0 + \sum_{i=1}^{z-1} d_i g_i = 0.   
\end{equation}
Now observe the pole structure of the functions $g_i$. Each $g_i$ has simple poles exactly at $P_i$ and $P_{i+1}$ (with the convention that if $\alpha_i = \alpha_{i+1}$, then $g_i = \frac{1}{X-\alpha_i}$ still has simple poles at both points because they share the same $x$-coordinate). No other $g_j$ has a pole at $P_1$ except $g_1$. Therefore, taking the principal part of \eqref{eq:linear_ind_proof} at $P_1$ isolates the term $d_1 g_1$, forcing $d_1 = 0$. Substituting $d_1 = 0$ back into \eqref{eq:linear_ind_proof}, we get $c_0 + \sum_{i=2}^{z-1} d_i g_i = 0$. At point $P_2$, the only function with a pole is now $g_2$ (since $g_1$ has been eliminated), so its principal part gives $d_2 = 0$. Repeating this argument successively for $P_3,\dots,P_{z-1}$, we obtain $d_1 = d_2 = \cdots = d_{z-1} = 0$. Finally, \eqref{eq:linear_ind_proof} reduces to $c_0 = 0$. Hence all coefficients are zero, proving linear independence.

By the Riemann–Roch theorem for elliptic curves:
\[
\dim \mathscr{L}(G) = \deg(G) = \sum_{i=1}^z k_i,
\]
\[
|\mathscr{L}_b| = 1 + \sum_{i=1}^z (k_i - 1) + (z - 1) = 1 + \left( \sum_{i=1}^z k_i - z \right) + (z - 1) = \sum_{i=1}^z k_i.
\]
As $\mathscr{L}_b$ consists of $\sum k_i$ linearly independent elements in $\mathscr{L}(G)$, it forms a basis.

\end{proof}
\end{theorem}

\begin{remark}
Clearly, to construct a basis for the Riemann–Roch space of the divisor $G = \sum_{i=1}^z k_i P_i$, we could use functions of the form:
\[
f_{s}(X,Y) = \delta_1\frac{Y + A_{s, 1}(X)}{(X - \alpha_1)^s}\cdot  \ldots \cdot \delta_z\frac{Y + A_{s, z}(X)}{(X - \alpha_z)^s}, \quad 2 \leq s \leq k, \delta_i \in \{0, 1\}.
\]

In Theorem~\ref{th:R-R_Arbitrary}, we specifically consider double-point basis functions to reduce the degree of basis elements. However, using Lemma~\ref{lem:R-R_single}, we can construct various different bases.
\\[2pt]
\end{remark}

\begin{remark}
We also note that Theorem \ref{th:R-R_Arbitrary} operates under the constraint $k_i \in \mathbb{N}_{\geq 2}$, which precludes its direct application for constructing the Riemann–Roch basis associated with divisors of the form $G = \sum_{i=1}^z k_i P_i$ when at least one $k_i = 1$.

Despite this, we can utilize the results of Lemma \ref{lem:R-R_single} and Theorem \ref{th:R-R_Arbitrary} to construct a basis in this case as well, by using not only pairs of consecutive points when constructing the functions $g_i(X,Y)$:
\[
g_{i,j}(X,Y) = \frac{Y + B_{i,j}(X)}{(X - \alpha_i)(X - \alpha_{j})}, \quad i \neq j.
\]

The drawback in this case is that the number of functions of the form $f_{i, s}(X, Y)$ and $g_{i,j}(X,Y)$ can vary significantly depending on the multiplicities of the points in the divisor $G$. The construction of a basis for arbitrary multiplicities is considered in Algorithm \ref{alg:R-R_arbitrary}.
\end{remark}

\begin{algorithm}[ht!]
\caption{Basis for arbitrary effective divisor $G = \sum_{i=1}^z k_i P_i$}
\label{alg:R-R_arbitrary}
\begin{algorithmic}[1]
\Require $\mathcal{E}/\FF_{q}:\mathcal{E}: Y^2 + a_1XY + a_3Y = X^3 + a_2X^2 + a_4X + a_6,$  

Points $P_i = P_{\alpha_i, \beta_i} = (\alpha_i, \beta_i) \in \mathcal{E}(\mathbb{F}_{p^m})$,

Integers $k_i \in \mathbb{N}_{\geq 1}$
\Ensure Basis $\mathscr{L}_b$ for $\mathscr{L}(G)$, where $G = \sum_{i=1}^z k_i P_i$.

\State $\mathscr{L}_b \gets \{1\}$ 
\State $k \gets \sum_{i=1}^z k_i$
\Statex //Basis functions with simple poles at $P_i$ and $P_j$//

\For{$i = 1$ \textbf{to} $z-1$}
\For{$j = i + 1$ \textbf{to} $z$}
\If{$P_i \neq - P_j$}
\State $\mathscr{L}_b \gets \mathscr{L}_b \cup \{\frac{Y + B_{i,j}(X)}{(X - \alpha_i)(X - \alpha_j)}\}$ \Comment{$B_{i,j}(X) = \left( \frac{\beta_j - \beta_i}{\alpha_j - \alpha_i} + a_1 \right) X + \left( (\beta_i + a_1\alpha_i + a_3) - \left( \frac{\beta_j - \beta_i}{\alpha_j - \alpha_i} + a_1 \right) \alpha_i \right)$}
\Else
\State $\mathscr{L}_b \gets \mathscr{L}_b \cup \{\frac{1}{X - \alpha_i}\}$
\EndIf
\If{$|\mathscr{L}_b| = k$} \State \Return $\mathscr{L}_b$ \EndIf
\EndFor
\EndFor

\Statex //Basis functions with poles at $P_i$//
\For{$i = 1$ \textbf{to} $z$}
\If{$P_i \not\in \mathcal{E}[2]$}
    \For{$s = 2$ \textbf{to} $k_i$}
        \State $\mathscr{L}_b \gets \mathscr{L}_b \cup \{\frac{Y + A_s(X)}{(X - \alpha_i)^s}\}$ \Comment{See Algorithm \ref{alg:R-R_single_char_2_3}}
        \If{$|\mathscr{L}_b| = k$} \State \Return $\mathscr{L}_b$ \EndIf
    \EndFor
\Else
    \Comment{$P_i$ is a 2-torsion point}
    \For{$s = 1$ \textbf{to} $\lfloor k_i/2 \rfloor$}
        \State $\mathscr{L}_b \gets \mathscr{L}_b \cup \{\frac{1}{(X - \alpha_i)^{s}}\}$ \Comment{For $s = 1$ check if $\frac{1}{(X - \alpha_i)} \not\in \mathscr{L}_b$}
        \If{$|\mathscr{L}_b| = k$} \State \Return $\mathscr{L}_b$ \EndIf
    \EndFor
    \For{$s = 2$ \textbf{to} $\lfloor (k_i+1)/2 \rfloor$}
        \State $\mathscr{L}_b \gets \mathscr{L}_b \cup \{\frac{Y - \beta_i}{(X - \alpha_i)^{s}}\}$
        \If{$|\mathscr{L}_b| = k$} \State \Return $\mathscr{L}_b$ \EndIf
    \EndFor
\EndIf
\EndFor
\end{algorithmic}
\end{algorithm}

\begin{remark}\label{rem:complexity_RR_arbitrary}
Algorithm~\ref{alg:R-R_arbitrary} constructs a basis for $\mathscr{L}(G)$ by first adding functions with simple poles at pairs of points (each computed in $\mathcal{O}(1)$ time) and then, for each point $P_i$, adding functions with higher order poles at $P_i$. The dominant part is the construction of the single-point basis for each $P_i \notin \mathcal{E}[2]$, which relies on Algorithm~\ref{alg:R-R_single_char_2_3}. For a fixed point $P_i$ with multiplicity $k_i$, the coefficients $c_j$ in the local expansion depend only on $P_i$ and can be precomputed once up to order $k_i$ at cost $\mathcal{O}(k_i^2)$. After that, each function $f_s$ (for $s=2,\dots,k_i$) can be constructed in $\mathcal{O}(1)$ time by iteratively updating the stored polynomial $A_s(X)$. For $P_i \in \mathcal{E}[2]$, the functions $\frac{1}{(X-\alpha_i)^s}$ and $\frac{Y-\beta_i}{(X-\alpha_i)^s}$ are also constructed in $\mathcal{O}(1)$ each. The construction of functions with simple poles at pairs of points requires $\mathcal{O}(z^2)$ operations. Hence the total complexity of the algorithm is $\mathcal{O}\!\left(z^2 + \sum_{i=1}^{z} k_i^2\right) \leq \mathcal{O}(k^2)$, where $k = \sum_{i=1}^{z} k_i$.
\end{remark}

\subsubsection{Bases for non-effective divisors}

We now extend the construction to divisors that include negative valuation coefficients. For a divisor  

\[
G = \sum_{i=1}^{z} k_i P_i \;-\; \sum_{j=1}^{r} \ell_j Q_j ,\qquad k_i,\ell_j\in\mathbb{N},
\]
with distinct points $P_i,Q_j\in\mathcal{E}(\mathbb{F}_{p^m})\setminus\{P_\infty\}$, the Riemann–Roch space is  

\[
\mathscr{L}(G)=\left\{f\in\mathbb{F}_q(\mathcal{E})\;:\; \operatorname{div}(f)+G\geq 0\right\}.
\]

The valuation conditions translate to:
\begin{itemize}
    \item at each $P_i$: $v_{P_i}(f)\geq -k_i$,
    \item at each $Q_j$: $v_{Q_j}(f)\geq \ell_j$,
    \item at every other point $R$ (including $P_\infty$): $v_R(f)\geq 0$.
\end{itemize}

If $\deg(G) = \sum k_i - \sum\ell_j < 0$, then $\mathscr{L}(G)=\{0\}$ by Riemann–Roch theorem. Hence we assume $\deg(G)\geq 0$ in the following. Let $G^+=\sum_{i=1}^z k_i P_i$  and $G^-=\sum_{j=1}^r \ell_j Q_j$. From the definition,
\begin{equation}\label{eq:non_effective_div}
  \mathscr{L}(G)=\bigl\{f\in\mathscr{L}(G^+)\;:\; v_{Q_j}(f)\geq \ell_j\ \text{for all }j\bigr\}.  
\end{equation}

Thus, once we have an explicit basis for $\mathscr{L}(G^+)$ (constructed by the methods of Theorem~\ref{th:R-R_Arbitrary} and Algorithm~\ref{alg:R-R_arbitrary}), the problem reduces to imposing a finite set of linear conditions on the coefficients of a function expressed in that basis.

Let $\mathscr{L}_b^+ = \{b_1,\dots,b_d\}$ be the basis of $\mathscr{L}(G^+)$ obtained from Theorem~\ref{th:R-R_Arbitrary}, where $d = \deg(G^+) = \sum_{i=1}^z k_i$. Any function $f\in\mathscr{L}(G^+)$ can be written uniquely as
\[
f = \sum_{\alpha=1}^{d} \lambda_\alpha b_\alpha,\qquad \lambda_\alpha\in\mathbb{F}_{p^m}.
\]

For a fixed negative point $Q_j = (\gamma_j,\delta_j)$ with multiplicity $\ell_j$, we use the local parameter $t_j = X - \gamma_j$. Expanding the elliptic curve equation around $Q_j$ gives
\[
Y = \delta_j + \sum_{m\geq 1} c_{j,m} t_j^m,
\]
where the coefficients $c_{j,m}$ are obtained recursively exactly as in Algorithm~\ref{alg:R-R_single_char_2_3}. Substituting this expansion into each basis function $b_\alpha$ (which is regular at $Q_j$ because $Q_j\notin\{P_i\}$) yields a Taylor series
\[
b_\alpha(X,Y) = \sum_{k\geq 0} b_{\alpha,j,k}\, t_j^k,\qquad b_{\alpha,j,k}\in\mathbb{F}_{p^m}.
\]

Hence
\[
f = \sum_{\alpha=1}^{d} \lambda_\alpha b_\alpha = \sum_{k\geq 0}\left(\sum_{\alpha=1}^{d} \lambda_\alpha b_{\alpha,j,k}\right) t_j^k.
\]

The condition $v_{Q_j}(f)\geq \ell_j$ is equivalent to the vanishing of the coefficients of $t_j^0,t_j^1,\dots,t_j^{\ell_j-1}$:
\begin{equation}\label{eq:local_conditions}
\sum_{\alpha=1}^{d} \lambda_\alpha b_{\alpha,j,k} = 0,\qquad k = 0,1,\dots,\ell_j-1.
\end{equation}

Collecting these equations for all points $Q_j$ gives a homogeneous linear system
\[
M \boldsymbol{\lambda} = \mathbf{0},\qquad \boldsymbol{\lambda}=(\lambda_1,\dots,\lambda_d)^T.
\]

The matrix $M$ has size $N\times d$ with $N = \sum_{j=1}^{r}\ell_j$. Its rows are indexed by pairs $(j,k)$ where $1\leq j\leq r$ and $0\leq k\leq \ell_j-1$. For such a pair, the corresponding row is
\[
\bigl(b_{1,j,k},\; b_{2,j,k},\; \dots,\; b_{d,j,k}\bigr).
\]

Equivalently, $\mathbf{M}$ has a block structure
\[
\mathbf{M} = \begin{pmatrix}
\mathbf{M}^{(1)} \\ \vdots \\ \mathbf{M}^{(r)}
\end{pmatrix},\qquad
\mathbf{M}^{(j)} = \begin{pmatrix}
b_{1,j,0} & \cdots & b_{d,j,0}\\[2pt]
b_{1,j,1} & \cdots & b_{d,j,1}\\[2pt]
\vdots & \ddots & \vdots\\[2pt]
b_{1,j,\ell_j-1} & \cdots & b_{d,j,\ell_j-1}
\end{pmatrix}\in\mathbb{F}_{p^m}^{\,\ell_j\times d}.
\]

The solution space of this system has dimension $d - N$, which by Riemann–Roch theorem equals $\deg(G)$. Any basis of the nullspace of $\mathbf{M}$ yields a basis of $\mathscr{L}(G)$.

\begin{theorem}\label{th:neg_basis}
Let $G = \sum_{i=1}^z k_i P_i - \sum_{j=1}^r \ell_j Q_j$ be a divisor on $\mathcal{E}/\mathbb{F}_{p^m}$ with $\deg(G)\geq 0$.  
Let $\mathscr{L}_b^+ = \{b_1,\dots,b_d\}$ be a basis of $\mathscr{L}(G^+)$ (obtained from Theorem~\ref{th:R-R_Arbitrary}), where $d = \deg(G)^+$.  
For each point $Q_j$ with negative valuation compute the Taylor expansion of every $b\in\mathscr{L}_b^+$ up to order $\ell_j-1$ using the local parameter $t_j = X-\gamma_j$.  
Collecting the coefficients as described yields a homogeneous linear system
\[
\mathbf{M} \boldsymbol{\lambda} = \mathbf{0},
\]
where $\mathbf{M}$ is an $N\times d$ matrix with $N=\sum_{j=1}^r \ell_j$.  
The solution space has dimension $d-N = \deg(G)$.  
A basis $\mathscr{L}_b$ of $\mathscr{L}(G)$ is obtained by taking a basis of the nullspace of $\mathbf{M}$ and forming the corresponding linear combinations of $\mathscr{L}_b^+$.
\end{theorem}

\begin{proof}
The reduction \eqref{eq:non_effective_div} shows that $\mathscr{L}(G)$ is exactly the subspace of $\mathscr{L}(G^+)$ consisting of functions that vanish to order at least $\ell_j$ at each $Q_j$. Each vanishing condition is linear in the coefficients $\lambda_\alpha$ because it involves the coefficients of a power series expansion. Riemann–Roch theorem gives $\dim\mathscr{L}(G)=\deg(G) = d-N$, therefore the nullspace of $\mathbf{M}$ has precisely this dimension, and any basis of it produces a basis of $\mathscr{L}(G)$.
\end{proof}

\begin{algorithm}[ht!]
\caption{Basis for $\mathscr{L}(G)$ with non-effective divisor $G$}
\label{alg:neg_basis}
\begin{algorithmic}[1]
\Require Elliptic curve $\mathcal{E}$ in general Weierstrass form, points $P_i=(\alpha_i,\beta_i)$, $Q_j=(\gamma_j,\delta_j)$ with $Q_j \notin \mathcal{E}[2]$, integers $k_i,\ell_j\geq 1$.
\Ensure Basis $\mathscr{L}_b$ for $\mathscr{L}(G)$ where $G=\sum k_iP_i-\sum\ell_jQ_j$.
\State Compute $d_{\text{pos}}=\sum k_i$, $d_{\text{neg}}=\sum\ell_j$. 
\If{$d_{\text{pos}}<d_{\text{neg}}$}
\State \Return $\varnothing$
\EndIf

\State Construct a basis $\mathscr{L}_b^+ = \{b_1,\dots,b_{d_{\text{pos}}}\}$ for $\mathscr{L}(G^+)$ using Algorithm~\ref{alg:R-R_arbitrary}.
\State $\mathbf{M}\leftarrow$ zero matrix of size $d_{\text{neg}}\times d_{\text{pos}}$.
\State $row\gets 1$
\For{each $Q_j$}
    \State Compute the local expansion of the curve at $Q_j$: determine coefficients $c_{j,m}$ in $Y = \delta_j + \sum_{m\ge1}c_{j,m} (X-\gamma_j)^m$ recursively \Comment{see Algorithm~\ref{alg:R-R_single_char_2_3}}.
    \For{$k=0$ to $\ell_j-1$}
        \For{$i=1$ to $d_{\text{pos}}$}
            \State Compute the coefficient of $(X-\gamma_j)^k$ in the expansion of $b_i$ by substituting the $Y$-expansion and extracting the term of degree $k$.
            \State Set $\mathbf{M}[row][i]$ to that coefficient.
        \EndFor
        \State $row\gets row+1$
    \EndFor
\EndFor
\State Compute a basis $\{\boldsymbol{\lambda}^{(1)},\dots,\boldsymbol{\lambda}^{(d_{\text{pos}}-d_{\text{neg}})}\}$ of the nullspace of $\mathbf{M}$.
\State For each $z=1,\dots,d_{\text{pos}}-d_{\text{neg}}$, define $f_z = \sum_{i=1}^{d_{\text{pos}}} \lambda_i^{(z)} b_i$.
\State $\mathscr{L}_b \gets \{f_1,\dots,f_{d_{\text{pos}}-d_{\text{neg}}}\}$
\State \Return $\mathscr{L}_b$.
\end{algorithmic}
\end{algorithm}

The basis functions obtained from Algorithm~\ref{alg:neg_basis} are explicit linear combinations of the elementary building blocks:
\[
f_{i,s}(X,Y) = \frac{Y + A_{i,s}(X)}{(X-\alpha_i)^s},\qquad 
g_{i,j}(X,Y) = \frac{Y + B_{i,j}(X)}{(X-\alpha_i)(X-\alpha_{j})}\ \ (\text{or } \frac{1}{X-\alpha_i}\text{ if }\alpha_i=\alpha_{j}),
\]
where $A_{i,s}$ and $B_{i,j}$ are polynomials defined in Lemma~\ref{lem:R-R_single} and Theorem~\ref{th:R-R_Arbitrary} (Algorithm \ref{alg:R-R_arbitrary}). Thus every $f\in\mathscr{L}(G)$ can be written uniquely as
\[
f = c_0 + \sum_{i=1}^z \sum_{s=2}^{k_i} c_{i,s} f_{i,s} + \sum_{i=1}^{z-1} d_i g_i.
\]

\begin{example}\label{ex:2P-Q}
Let $\mathcal{E}: Y^2 = X^3 + a_4X + a_6$ over a field of characteristic $>3$. Consider $G = 2P - Q$ with $P=(\alpha,\beta)$, $Q=(\gamma,\delta)$, and assume $P\neq Q$ and $P\neq -Q$. A basis for $\mathscr{L}(2P)$ is $\{1, f_2\}$ where $f_2 = \frac{Y + A_2(X)}{(X-\alpha)^2}$ with $A_2(X) = -\beta - c_1(X-\alpha)$ and $c_1 = \frac{3\alpha^2+a_4}{2\beta}$. To find a function in $\mathscr{L}(2P-Q)$, we need a linear combination $c_0 + c_2 f_2$ that vanishes at $Q$. The condition $c_0 + c_2 f_2(Q)=0$ gives one linear relation. Thus a basis is $\{f_2 - f_2(Q)\cdot 1\}$. Explicitly,
\[
f = \frac{Y + A_2(X)}{(X-\alpha)^2} - \frac{\delta + A_2(\gamma)}{(\gamma-\alpha)^2}.
\]
This function has a double pole at $P$ and a simple zero at $Q$.
\end{example}

\begin{remark}
When the divisor has only one positive point and one negative point, i.e., $G = kP - \ell Q$ with $P=(\alpha,\beta)$, $Q=(\gamma,\delta)$ and $k>\ell$, the construction simplifies. The basis of $\mathscr{L}(kP)$ is $\mathscr{L}_b^+ = \{1, f_2, \dots, f_k\}$ where
\[
f_s(X,Y) = \frac{Y + A_s(X)}{(X-\alpha)^s},\qquad 2\leq s\leq k,
\]
as given in Lemma~\ref{lem:R-R_single}.  
Using the local parameter $t = X-\gamma$ at $Q$, each $f_s$ expands as
\[
f_s = f_s(Q) + f_s'(Q)\,t + \frac{1}{2}f_s''(Q)\,t^2 + \cdots,
\]
where the derivatives are taken with respect to $t$.  
For a linear combination $f = c_0 + \sum_{s=2}^k c_s f_s$, the conditions $v_Q(f)\geq \ell$ become
\[
c_0 + \sum_{s=2}^k c_s f_s(Q) = 0,\quad
\sum_{s=2}^k c_s f_s'(Q) = 0,\quad \dots,\quad
\sum_{s=2}^k c_s f_s^{(\ell-1)}(Q) = 0.
\]
This is a homogeneous linear system of $\ell$ equations in the $k$ unknowns $c_0,\dots,c_k$. Its solution space has dimension $k-\ell$, matching $\deg(G)$.  
For $\ell=1$ we recover the simple condition $c_0 + \sum_{s=2}^k c_s f_s(Q)=0$; in the case $k=2$ this gives the basis element $f_2 - f_2(Q)\cdot 1$ (as in Example~\ref{ex:2P-Q}). For $\ell=2$ one must also enforce the vanishing of the first derivative.
\end{remark}

\begin{example}[$k=3$, $\ell=2$]\label{ex:3P-2Q}
As in Example \ref{ex:2P-Q}, assume $\operatorname{char}\neq 2,3$ and take $G = 3P - 2Q$ with $P=(\alpha,\beta)$, $Q=(\gamma,\delta)$ and $\alpha\neq\gamma$.  
A basis of $\mathscr{L}(3P)$ is $\{1,f_2,f_3\}$ where
\[
f_2=\frac{Y+A_2(X)}{(X-\alpha)^2},\quad 
f_3=\frac{Y+A_3(X)}{(X-\alpha)^3},
\]
and $A_2$, $A_3$ are constructed as in Lemma~\ref{lem:R-R_single}.  
We need $c_0,c_2,c_3$ such that $f(Q)=0$ and $f'(Q)=0$. Expanding $f_2$, $f_3$ at $Q$ up to order $1$ gives
\[
f_2 = f_2(Q) + f_2'(Q) t + \mathcal{O}(t^2),\quad
f_3 = f_3(Q) + f_3'(Q) t + \mathcal{O}(t^2).
\]
The conditions become
\[
c_0 + c_2 f_2(Q) + c_3 f_3(Q) = 0,\qquad
c_2 f_2'(Q) + c_3 f_3'(Q) = 0.
\]
Assuming the coefficient matrix has rank $2$, we obtain a one‑dimensional solution space.  Solving, e.g., taking $c_3=1$ gives
\[
c_2 = -\frac{f_3'(Q)}{f_2'(Q)},\qquad
c_0 = -f_2(Q)c_2 - f_3(Q).
\]
Thus
\[
f = -\bigl(f_2(Q)c_2 + f_3(Q)\bigr) + c_2 f_2 + f_3
\]
has a zero of order 2 at $Q$ and a triple pole at $P$.
\end{example}

The construction in Example \ref{ex:3P-2Q} is a special case of the general method described in Theorem~\ref{th:neg_basis} and Algorithm~\ref{alg:neg_basis}.  For divisors with several points, one simply enlarges the linear system to include conditions at each point with negative valuation coefficient.

\subsection{Riemann–Roch spaces for QC-Elliptic codes and their subfield subcodes}\label{sec:QC}

In this section we will consider Riemann–Roch spaces associated with divisors involved in the construction of Quasi-cyclic Elliptic Codes and their subfield subcodes. The general construction of Subfield Subcodes of Algebraic Geometry codes (SSAG) is described in~\cite{Bar}. We will show that knowledge of the automorphisms action significantly simplifies the construction of QC-elliptic codes compared to standard elliptic codes related to arbitrary divisors $G \neq kP_\infty$. Quasi-cyclic Subfield Subcodes of Dual Elliptic (QC-SSDE) codes are particularly interesting for applications in code-based cryptosystems due to the possibility of defining such codes with fewer vectors. Thanks to the quasi-cyclicity property, when embedding this family of codes into the McEliece type cryptosystem~\cite{McEliece78}, it becomes possible to achieve significant reduction in key sizes compared to the Classic McEliece~\cite{ClassicMcEliece} based on binary Goppa codes.

\begin{definition}[\textit{Quasi-Cyclic Code}]\label{Def:QC}
    Let $\mathit{n}$ be an integer and $\ell \mid \mathnormal{n}$. A code $\mathcal{C} \subseteq  \mathbb{F}^{n}$ is called \textit{$\ell$-quasi-cyclic} ($\mathit{\ell}$-QC) if for every $c \in \mathcal{C}$, $\sigma(c) \in \mathcal{C}$, where $\sigma$ is the quasi-cyclic shift
    \[ 
    \;
    \sigma:
    \begin{cases}
        \mathbb{F}^\mathnormal{n} \to \mathbb{F}^\mathnormal{n}, \\
        \left({\bf b}_1 | \ldots | {\bf b}_{\frac{\mathnormal{n}}{\ell}}\right) \mapsto \left(\sigma_{\ell}({\bf b}_1) | \ldots | \sigma_{\ell} ({\bf b}_{\frac{n}{\ell}})\right).
    \end{cases}
    \] 
    Here, $\sigma_\ell$ is cyclic shift of blocks $b_i$ of length $\ell$ and $\ell$ is called the \textit{order of quasi-cyclicity} of the code.
\end{definition}

The primary advantage of quasi-cyclic codes is the smaller public key size, which results from defining the matrix with fewer vectors. Similar to cyclic codes, which are defined by circulant matrices, a quasi-cyclic code can be defined using a so-called \textit{block-circulant matrix}.
\begin{definition}[Block-Circulant Matrix]
    A matrix $\mathbf{M}$ is called \textit{$\ell$-block-circulant} if it consists of $\mathbf{M}_{\mathit{i}}$ circulant matrices of order $\ell$:
    \begingroup
\renewcommand*{\arraystretch}{1.2}
\[
\mathbf{M} = 
\begin{pmatrix}
\mathrm{rot}(\mathbf{a}_{1,1}) & \mathrm{rot}(\mathbf{a}_{1,2}) & \cdots & \mathrm{rot}(\mathbf{a}_{1,n/\ell}) \\
\vdots & \vdots & \ddots & \vdots \\
\mathrm{rot}(\mathbf{a}_{k/\ell,1}) & \mathrm{rot}(\mathbf{a}_{k/\ell,2})& \cdots & \mathrm{rot}(\mathbf{a}_{k/\ell,n/\ell}) 
\end{pmatrix} \in \FF^{k \times n}, \text{ where } \mathrm{rot}(\mathbf{a}_{i,j}) = 
\begin{pmatrix}
a^{(i,j)}_0 & a^{(i,j)}_1 & \dots & a^{(i,j)}_{ \ell-1} \\
a^{(i,j)}_{ \ell-1}& a^{(i,j)}_0 & \dots & a^{(i,j)}_{ \ell-2} \\
\vdots & \ddots & \ddots & \vdots \\
a^{(i,j)}_1 & \dots & a^{(i,j)}_{ \ell-1} & a^{(i,j)}_0 \\
\end{pmatrix}.
\]
\endgroup
\end{definition}

We now consider a definition of quasi-cyclic elliptic code defined over $\mathbb{F}_{q}$ following the notation of \cite{Bar}. 
Let $\sigma \in \operatorname{Aut}(\mathcal{X}/\FF_q)$ with $\operatorname{ord}(\sigma) = \ell$. For a rational point $P \in \mathcal{X}(\FF_q)$, define its orbit as:
\[
\operatorname{Orb}_{\sigma}(P) = \left\{\sigma^{i}(P) : i \in \{0, \ldots, \ell-1\}\right\}.
\]

\begin{definition}[QC-elliptic code]\label{def:QC-Elliptic}
Let $\mathcal{E}/\FF_{p^m}$ be an elliptic curve and $\sigma \in \operatorname{Aut}(\mathcal{E}/\FF_{p^m})$ be an automorphism of order $\operatorname{ord}(\sigma)=\ell$. The elliptic code $\cC_{\mathscr{L}}(D, G)$ is called $\ell$-quasi-cyclic if the  divisors $D$ and $G$ are of the form:
\begin{equation}\label{eq:QC_Divisors}
\begin{gathered}
D = \textstyle\sum\limits_{i=1}^{n/\ell} \sum\limits_{P \in \operatorname{Orb}_{\sigma}(P_{i})} P, \quad \text{supp}(D)=\coprod\limits_{i=1}^{n / \ell} \operatorname{Orb}_{\sigma}(P_{i}),\\
G=\textstyle\sum\limits_ {i=1}^{s} t_{i} \sum\limits_{Q \in \operatorname{Orb}_{\sigma}(Q_{i})} Q,
\end{gathered}
\end{equation}

where $P_{i}, Q_{i} \in \mathcal{E}(\FF_q)$ are pairwise distinct points that are not invariant under the given mapping $\sigma$, with disjoint orbits of size $\ell$, and $\operatorname{Supp}(D) \cap \operatorname{Supp}(G) = \varnothing$, $s \in \mathbb{N}$, $t_{i} \in \ZZ$ for $i \in\{1, \ldots, s\}$.
\end{definition}

Next, we will demonstrate that for quasi-cyclic elliptic codes, a more efficient construction of the Riemann–Roch space basis is possible. While constructing QC-SSDE codes, the orbit of each point $P \in \operatorname{Supp}(D)$ must contain $\ell$ distinct elements. In~\cite{TD13}, it was proven that the action of $\sigma$ on all rational points of $\mathcal{E}/\FF_q$ gives rise to orbits, each containing $\operatorname{ord} \sigma$ elements, except possibly for orbits that include the points $P_{\alpha, \beta} \in \mathcal{E}[2] \cup \mathcal{E}[3] \cup \{P_{0, \beta_1}, P_{0, \beta_2}\}$. By examining the action of automorphisms on the points of an elliptic curve, in Theorem~\ref{th:R-R-QC} we obtain a simpler form of the Riemann–Roch space basis in the QC-case.

\begin{theorem}\label{th:R-R-QC}
    Let $\mathcal{E}/\FF_{p^m}$ be an elliptic curve, $p \geq 2$, $\sigma \in \operatorname{Aut}(\mathcal{E}/\FF_{p^m}, P_\infty)$, $\ell = \operatorname{ord}(\sigma) \in \{2, 4, 6\}$, and $G = \textstyle\sum\limits_{z=1}^{s} t_{z} \sum\limits_{Q \in \operatorname{Orb}_{\sigma}(Q_{z})} Q$, where $s \in \mathbb{N}$ and $Q_z \in \mathcal{E}(\FF_{p^m})\backslash (\{P_\infty, P_{0, \beta_1}, P_{0, \beta_2}, P_{\alpha_1, 0}, P_{\alpha_2, 0}, P_{\alpha_3, 0}\} \cup \mathcal{E}[3])$. Also, denote $\mathcal{Q}_z = \operatorname{Orb}_{\sigma}(Q_{z})$, and let $\mathcal{Q}_{z,\rho}$ be the $\rho$-th element of this orbit for $\rho = 1,\dots,\ell$. The basis of the Riemann–Roch space associated with the divisor $G$ is defined as follows:
    \begin{equation}\label{eq:R-R-basis-QC}
        \mathscr{L}_{b}(G) = \left\{\frac{X^iY^j}{\textstyle\prod_{z = 1}^{s} \left[\textstyle\prod_{\rho = 1}^{\ell/2}\left(X-x(\mathcal{Q}_{z,\rho})\right)\right]^{t_z}} \bigg| 2i+3j \leq \ell\cdot \textstyle\sum_{z = 1}^{s} t_z,j = \overline{0, 1} \right\},
    \end{equation}
    where $x(\mathcal{Q}_{z,\rho})$ is the $x$-coordinate of the point $\mathcal{Q}_{z,\rho}$ and $\deg(G) = \ell\cdot \textstyle\sum_{z = 1}^{s} t_z$.

\begin{proof}
    Since the linear independence of the considered functions is obvious, it remains to show that functions of this form indeed belong to the Riemann–Roch space
    \[
    \mathscr{L}(G) = \{f \in \FF_q(\cX) : (f) \geq -G \}.
    \]
    
    First, consider the action of automorphisms $\sigma \in \operatorname{Aut}(E/\FF_{p^m}, P_\infty)$ on the points of the curve in the case where $\operatorname{ord}(\sigma) \in \{2, 4, 6\}$ (see Table~\ref{QC:Table_sigma}).
\begin{table}[ht!]
\centering
\begin{tabular}{cc}
				\hline
				\multicolumn{2}{c}{$\operatorname{char}(\FF_q) = 2$}\\
				\hline
				$\operatorname{ord}(\sigma)=4\Rightarrow u = 1$ & $\operatorname{ord}(\sigma)=6\Rightarrow u^2=u+1$ \\
				
				$\sigma(x)=x+r,\quad \sigma(y)=y+sx+t$ & $\sigma(x)=ux+x+r,\quad \sigma(y)=y+usx+sx+t$ \\
				$\sigma^2(x)=x,\quad \sigma^2(y)=y+sr$ & $\sigma^2(x)=ux+ur,\quad \sigma^2(y)=y+sx+usr+sr$ \\
				$\sigma^3(x)=x+r,\quad \sigma^3(y)=y+sr+t$ & $\sigma^3(x)=x,\quad \sigma^3(y)=y+sx+usr+t$ \\
				$\sigma^4(x)=x,\quad \sigma^4(y) = y$ & $\sigma^4(x)=ux+x+r,\quad \sigma^4(y)=y+usx+sx+usr$ \\
				& $\sigma^5(x)=ux+ur,\quad \sigma^5(y)=y+sx+sr+t$ \\
				\hline
				\multicolumn{2}{c}{$\operatorname{char}(\FF_q) = 3$}\\
				\hline
				$\operatorname{ord}(\sigma)=4\Rightarrow u^4=1$ & $\operatorname{ord}(\sigma)=6\Rightarrow u^2=1$ \\
				
				$\sigma(x)=-x+r,\quad \sigma(y)=u^3y$ & $\sigma(x)=x+r,\quad \sigma(y)=uy$ \\
				$\sigma^2(x)=x,\quad \sigma^2(y)=-y$ & $\sigma^2(x)=x+2r,\quad \sigma^2(y)=y$ \\
				$\sigma^3(x)=-x+r,\quad \sigma^3(y)=uy$ & $\sigma^3(x)=x,\quad \sigma^3(y)=uy$ \\
				$\sigma^4(x)=x,\quad \sigma^4(y) = y$ & $\sigma^4(x)=x+r,\quad \sigma^4(y)=y$ \\
				& $\sigma^5(x)=x+2r,\quad \sigma^5(y)=uy$ \\
				\hline
				\multicolumn{2}{c}{$\operatorname{char}(\FF_q) \geq 3$}\\
				\hline
				$\operatorname{ord}(\sigma)=4\Rightarrow u^4=1$ & $\operatorname{ord}(\sigma)=6\Rightarrow u^6=1$ \\
				
				$\sigma(x)=u^2x,\quad \sigma(y)=u^3y$ & $\sigma(x)=u^2x,\quad \sigma(y)=u^3y$ \\
				$\sigma^2(x)=x,\quad \sigma^2(y)=u^2y$ & $\sigma^2(x)=u^4x,\quad \sigma^2(y)=y$ \\
				$\sigma^3(x)=u^2x,\quad \sigma^3(y)=uy$ & $\sigma^3(x)=x,\quad \sigma^3(y)=u^3y$ \\
				$\sigma^4(x)=x,\quad \sigma^4(y) = y$ & $\sigma^4(x)=u^2x,\quad \sigma^4(y)=y$ \\
				& $\sigma^5(x)=u^4x,\quad \sigma^5(y)=u^3y$ \\
				
				\hline
				\multicolumn{2}{c}{$\operatorname{ord}(\sigma)=2 \Rightarrow u = -1, \quad \sigma(x)=x, \quad \sigma(y)=-y$}\\
				\hline
			\end{tabular}
\caption{Action of automorphisms on elliptic curve points}\label{QC:Table_sigma}%
\end{table}    
Based on the action of automorphisms on the points of the elliptic curve, the divisor $G$ can always be represented in the following form: 
\[
G = \textstyle\sum\limits_{z=1}^{s} t_{z}(\mathcal{Q}_{z,1} + \dots + \mathcal{Q}_{z,\ell}), \text{ where } x(\mathcal{Q}_{z,\rho}) = x(\mathcal{Q}_{z,\rho+\ell/2}) \text{ for } \rho \in \{1, \ell/2\}.
\]
Note that all points with the same $x$-coordinate always lie in the same orbit $\mathcal{Q}_{z}$. We use this fact when constructing the basis of the Riemann–Roch space. Consider the divisors of the function 
\[
f = \frac{X^iY^j}{\textstyle\prod_{z = 1}^{s} \left[\textstyle\prod_{\rho = 1}^{\ell/2}\left(X-x(\mathcal{Q}_{z,\rho})\right)\right]^{t_z}}.
\]
Clearly, the poles of $f$ can only be the points $\mathcal{Q}_{z,\rho}$ and possibly the point at infinity.

Taking into account that $v_P(XY)=v_P(X)+v_P(Y)$ and $v_{P_\infty}(X^iY^j) = iv_{P_\infty}(X) + jv_{P_\infty}(Y) = -2i - 3j$, we obtain:
\begin{equation*}
		\begin{split}
			&\left(f\right)_0 = \quad \sum_{\mathclap{P \in \mathcal{Z}(X^iY^j) \backslash P_\infty}}v_{P}(X^iY^j)P +(\sum_{z=1}^{s}\ell t_z)P_\infty,\\
			&\left(f\right)_\infty = \sum\limits_{z=1}^{s} t_{z}(\mathcal{Q}_{z,1} + \dots + \mathcal{Q}_{z,\ell}) + (2i+3j)P_\infty= G + (2i+3j)P_\infty,\\
			&\left(f\right) = \quad  \sum_{\mathclap{P \in \mathcal{Z}(X^iY^j) \backslash P_\infty}}v_{P}(X^iY^j)P +(\sum_{z=1}^{s}\ell t_z)P_\infty  - G - (2i+3j)P_\infty.
		\end{split}
	\end{equation*}

Since $\mathcal{Q}_{z,\rho} \in \mathcal{E}(\FF_{p^m})\backslash (\{P_\infty, P_{0, \beta_1}, P_{0, \beta_2}, P_{\alpha_1, 0}, P_{\alpha_2, 0}, P_{\alpha_3, 0}\} \cup \mathcal{E}[3])$, we conclude that $(f) \geq -G$ when $2i+3j \leq \ell\sum_{z=1}^{s}t_z$. 
\end{proof}
\end{theorem}

\begin{remark}
Note that Theorem~\ref{th:R-R-QC} is not applicable in the case of order $3$ automorphisms, since in this case points with the same $x$-coordinates belong to different orbits. However, thanks to Theorem~\ref{th:R-R_Arbitrary}, an effective construction of the basis is possible in this case as well.
\end{remark}

\begin{corollary}\label{cor:RR-neg-inf}
    Under the same assumptions as Theorem~\ref{th:R-R-QC}, for the divisor $G = \sum_{z=1}^{s} t_{z} \sum_{Q \in \operatorname{Orb}_{\sigma}(Q_{z})} Q - c P_\infty$ with $c \leq \ell\sum_{z=1}^{s} t_z$, the Riemann–Roch space $\mathscr{L}(G)$ has a basis:
   \begin{equation*}
        \mathscr{L}_{b}(G) = \left\{\frac{X^iY^j}{\textstyle\prod_{z = 1}^{s} \left[\textstyle\prod_{\rho = 1}^{\ell/2}\left(X-x(\mathcal{Q}_{z,\rho})\right)\right]^{t_z}} \bigg| 2i+3j \leq \ell\cdot \sum_{z=1}^{s} t_z - c,j = \overline{0, 1} \right\},
    \end{equation*}
    where $\deg(G) = \ell\sum_{z=1}^{s} t_z - c$. 

\begin{proof}
    Let $G_0 = \sum_{z=1}^{s} t_{z} \sum_{Q \in \operatorname{Orb}_{\sigma}(Q_{z})} Q$, so that $G = G_0 - c P_\infty$ and $\deg(G_0) = \ell \sum_{z=1}^{s} t_z$. Consider the function
    \[
    f = \frac{X^iY^j}{\prod_{z = 1}^{s} \left[\prod_{\rho = 1}^{\ell/2}\left(X-x(\mathcal{Q}_{z,\rho})\right)\right]^{t_z}}.
    \]
    As established in the proof of Theorem~\ref{th:R-R-QC}, the divisor of $f$ is
    \[
    \operatorname{div}(f) = (X^iY^j)_0 - G_0 + \left(\deg(G_0) - 2i - 3j\right) P_\infty,
    \]

    Thus, $v_{P_\infty}(f) \geq c$ if and only if
    \[
    \deg(G_0) - 2i - 3j \geq c \iff 2i + 3j \leq \deg(G_0) - c.
    \]
    Given that $\deg(G_0) - c = \deg(G)$, this condition is $2i + 3j \leq \deg(G)$. 
\end{proof}
\end{corollary}

Next, we will examine bases and the number of points appearing in the support depending on the field characteristic. Indeed, different characteristics may induce different automorphisms and, consequently, distinct orbit sizes for the same points. The remaining case to address concerns divisors supported on invariant points or on points whose orbit cardinality is strictly smaller than the order of the automorphism. Such divisors are of the form
\[
G = \sum_{z=1}^{s} t_{z} \sum_{Q \in \operatorname{Orb}_{\sigma}(Q_{z})} Q, \text{ where }  s \in \mathbb{N}  \text{ and } Q_z \in \{P_{0, \beta_1}, P_{0, \beta_2}, P_{\alpha_1, 0}, P_{\alpha_2, 0}, P_{\alpha_3, 0}\} \cup \mathcal{E}[3].
\]

Table~\ref{tab:invariant_points_orbits} lists the orbit sizes corresponding to the possible invariant points $ Q_z $. This information allows to explicitly construct bases of the associated Riemann–Roch spaces for divisors containing one or more invariant points.

\begin{table}[ht!]
    \centering
    \begin{tabular}{cccc}
        \hline
        $\operatorname{char}(\mathbb{F}_q)$ & $\ell = \operatorname{ord}(\sigma)$ & Invariant Point(s) $P$ & $\omega_\sigma(P) = |\operatorname{Orb}_\sigma(P)|$ \\
        \hline
        
        \multirow{4}{*}{$p > 3$} & 2,4 & $P_{\alpha_i, 0}$ & $\begin{cases} 1 & (\ell=2) \\ 2 & (\ell=4) \end{cases}$ \\
        \cline{2-4}
        & 3 & $P_{0,\beta_j}$ & $1$ \\
        \cline{2-4}
        & \multirow{2}{*}{6} & $P_{0,\beta_j}$ & $2$ \\
        \cline{3-4}
        & & $P_{\alpha_i, 0}$ & $3$ \\
        \hline

        \multirow{4}{*}{$p = 3$} & 2,4,6 & $P_{\alpha_i,0}$ & $\begin{cases} 1 & (\ell=2) \\ 2 & (\ell=4) \\ 3 & (\ell=6)\end{cases}$ \\
        \cline{2-4}
        & 3 & $-$ & $-$ \\
        
        \hline\\
        \multirow{4}{*}{$p = 2$} & 2 & $ j \neq 0, P_{\alpha_i,0}$ & $1$ \\
        \cline{2-4}
        & 3 & $P \in \mathcal{E}[3]$ & $1$ \\
        \cline{2-4}
        & 4 & $-$ & $-$  \\
        \cline{2-4}
        & 6 & $P \in \mathcal{E}[3]$ & $1$ \\
        \hline
    \end{tabular}
    \caption{Orbit sizes $\omega_\sigma(P)$ of invariant points under an automorphism $\sigma$ of order $\ell$}
    \label{tab:invariant_points_orbits}
\end{table}

\begin{remark}
Although the classical construction of QC codes considers supports that do not include invariant points, adding one or several such points to the supports of the divisors $D$ and $G$ can significantly increase the variability of code parameters, since in this case the constraints $\ell \mid n$ and $\ell \mid k$  can be omitted without losing the possibility of constructing a block-circulant generator matrix. We also note that invariant points are convenient to use when constructing MDS and iso-dual codes, which we consider in Section \ref{sec:mds_iso-dual}.
\end{remark}

\paragraph{Cryptographic application of Quasi-cyclic Subfield Subcodes of Dual Elliptic codes (QC-SSDE)}
We emphasize that the McEliece cryptosystem based on algebraic geometry codes has been effectively broken for nearly all parameter choices in~\cite{CMP17}. Currently, there exists only one scheme, $\text{ECC}^2$~\cite{ECC2}, utilizing elliptic codes that remains resistant to the attack described in~\cite{CMP17}. However, the overall security of this scheme, as well as its practical applicability, remains questionable. Since quasi-cyclic codes are of particular interest for cryptographic applications, we focus on elliptic code subcodes that demonstrate resistance against all currently known attacks. Specifically, we examine the family of quasi-cyclic subfield subcodes of elliptic codes.

\begin{definition}[Subfield subcode]
    Let $\mathcal{C} \subset \mathbb{F}_{q^{m}}^{n}$ be a linear $[n, k, d]$ code over $\mathbb{F}_{q^{m}}$. The \textit{subfield subcode} of $\mathcal{C}$ over $\mathbb{F}_{q}$ is defined as $\left.\mathcal{C}\right|_{\mathbb{F}_{q}}$, comprising codewords whose components lie in $\mathbb{F}_{q}$:
    \[
    \left.\mathcal{C}\right|_{\mathbb{F}_{q}} = \mathcal{C} \cap \mathbb{F}_{q}^{n}.
    \]   

   $\left.\mathcal{C}\right|_{\mathbb{F}_{q}}$ is an $\left[n, \tilde{k}, \tilde{d}\right]$-code over $\mathbb{F}_{q}$ satisfying:
    \begin{equation}\label{Sub:params_base}
        \tilde{k} \geq n - m(n - k) \quad \text{and} \quad \tilde{d} \geq d.
    \end{equation}
\end{definition}

\begin{remark} Note that if elliptic code $\mathcal{C}$ is associated with divisors $D, G$ of the form \eqref{eq:QC_Divisors}, then its dual code $\cC^\perp$ and subfield subcodes $\left.\cC\right|_{\mathbb{F}_{p}}$ are also quasi-cyclic. Thus, we can define $\mathcal{C}_{\operatorname{QC-SSDE}}(D, G) = \left.\cC_{\mathscr{L}}(D, G)^\perp\right|_{\mathbb{F}_{p}}$, associated with divisors \eqref{eq:QC_Divisors}.
\end{remark}

\begin{algorithm}[H]
\caption{Construction of $\operatorname{QC-SSDE}$ Code (Adapted from~\cite{BIG_QUAKE})}
\label{Alg:SSE_theory}
\begin{algorithmic}[1]
\Require Field $\FF_{p^m}$, elliptic curve $\mathcal{E}/\FF_{p^m}$, parameters $[n, k^\star ]$ of the original code $\cC_{\mathscr{L}}(D, G)$, quasi-cyclic order $\ell$, automorphism $\sigma_\ell \in \operatorname{Aut}(\mathcal{E}/\FF_{p^m}, P_\infty)$.
\Ensure Parity-check matrix $\mathbf{H}_{\operatorname{QC-SSDE}}$ of the code $\mathcal{C}_{\operatorname{QC-SSDE}}(D, G) = \left.\cC_{\mathscr{L}}(D, G)^\perp\right|_{\mathbb{F}_{p}}$.
\State $\mathscr{L}_b(G) \gets \begin{cases}
x^iy^j \mid 2i+3j \leq k^\star ,j = \overline{0, 1}& \text{if } G = k^\star P_\infty\\
\dfrac{X^iY^j}{\textstyle\prod_{z = 1}^{s} \left[\textstyle\prod_{\rho = 1}^{\ell/2}\left(X-x(\mathcal{Q}_{z,\rho})\right)\right]^{t_z}} \bigg| 2i+3j \leq \ell\cdot \sum_{z=1}^{s} t_z = k^\star & \text{if } G = \textstyle\sum\limits_{z=1}^{s} t_{z} \sum\limits_{Q \in \operatorname{Orb}_{\sigma}(Q_{z})} Q
\end{cases}$
\State $P_0, \dots, P_{\frac{n}{\ell}-1} \gets$ random points of $\mathcal{E}(\FF_{p^m})$ such that for each pair $i, j \in \{0, \dots, \frac{n}{\ell}-1\}$, where $i\neq j$, and for any $s \in \{0, \dots, \ell-1\}$, it holds that $P_i\neq\sigma_{\ell}^{s}(P_j)$
\State $D_{\operatorname{Supp}} \gets \left(P_0, \sigma_\ell(P_0), \sigma_{\ell}^{2}(P_0), \dots, \sigma_{\ell}^{\ell-1}(P_0), P_1, \sigma(P_1), \dots, \sigma_{\ell}^{\ell-1}(P_{\frac{n}{\ell}-1})\right)$
\State $D \gets \sum\limits_{0 \leq i \leq n-1} D_{\operatorname{Supp}}[i]$
\State $H_0 \gets $ parity-check matrix of the code $\cC_\mathscr{L}(D, G)^\perp$ \Comment{generator matrix of the code $\cC_\mathscr{L}(D, G)$}
\State $\widecheck{h}_{j i} \gets \left(h_{j i}^{(1)}, h_{j i}^{(2)}, \ldots, h_{j i}^{(m)}\right)$ \Comment{decomposition of ${h}_{j i} \in H_0$ over $\FF$ basis}
\State $\widecheck{H}_{0} \gets \{\widecheck{h}^{T}_{ji}\}$
\State $\widecheck{H}_{0} \gets$ apply the Gaussian elimination method to $\widecheck{H}_{0}$, remove zero rows
\State $\mathbf{H}_{\operatorname{QC-SSDE}} \gets \widecheck{H}_{0}^{(n-k) \times n}$
\State \Return $\mathbf{H}_{\operatorname{QC-SSDE}}$.
\end{algorithmic}
\end{algorithm}

For the code parameters currently used in the McEliece cryptosystem, no polynomial‑time structural attacks are known against QC‑SSDE codes. Moreover, by considering the automorphism group structure of elliptic curves (see Theorems~\ref{Aut_EC:Aut_theorem2} and~\ref{Aut_EC:Aut_remark}), we can achieve a $\ell$-fold reduction in public key size, where $\ell \in \{2, 3, 4, 6\}$. These properties make QC-SSDE codes particularly promising for cryptographic applications. Moreover, by applying Theorem~\ref{th:R-R-QC} along with Corollary~\ref{cor:RR-neg-inf}, we can effectively construct QC-SSDE codes with arbitrary divisors. This significantly increases parameter variability and improves potential scheme's security.

\paragraph{Construction of quasi-cyclic codes using arbitrary automorphisms}

A fundamental drawback in the construction of quasi-cyclic elliptic codes associated with divisors of the form $G = kP_\infty$ or $G = \textstyle\sum\limits_{z=1}^{s} t_{z} \sum\limits_{Q \in \operatorname{Orb}_{\sigma}(Q_{z})} Q$ with $\sigma \in \operatorname{Aut}(\mathcal{E}/\mathbb{F}_{q}, P_\infty)$ is the restriction on the order of the automorphisms. It is known (see Theorem \ref{Aut_EC:Aut_theorem2}) that the group $\operatorname{Aut}(\mathcal{E}/\mathbb{F}_{q}, P_\infty)$ has a cyclic subgroup of order not exceeding $\ell = 6$. Thus, using divisors of this form, it is impossible to construct QC codes with a larger quasi-cyclic order.

Let $ \mathbb{P}_E^{(1)} $ be the set of rational points of the elliptic function field $ E/\mathbb{F}_q $. It is known that function field of any algebraic curve is isomorphic to some algebraic function field $F/\FF_q$ of the same genus. Consider the action of the elements of the automorphism group $ \operatorname{Aut}(E/\mathbb{F}_q) $ on the rational points of the function field. Let $ \sigma \in \operatorname{Aut}(E/\mathbb{F}_q) $ and $ P \in \mathbb{P}_E^{(1)} $. Then $ \sigma(P) $ is also a point of degree one. The following theorem gives the structure of the group $\operatorname{Aut}(E/\mathbb{F}_q)$.

\begin{theorem}[{\cite[Theorem 3.1]{MX20}}]\label{th:aut}
Let $ E $ be an elliptic function field defined over $ \mathbb{F}_q $. Denote by $ \operatorname{Aut}(E/\mathbb{F}_q, P_\infty) $ the stabilizer of the point at infinity and by $ T_E = \{ \tau_Q : Q \in \mathbb{P}_E^{(1)} \} $ the translation group of the function field $ E $ over $ \mathbb{F}_q $. Then the automorphism group of the elliptic function field over $ \mathbb{F}_q $ is a semidirect product of the translation group $ T_E $ and the stabilizer $ \operatorname{Aut}(E/\mathbb{F}_q, P_\infty) $, i.e.,
\[
\operatorname{Aut}(E/\mathbb{F}_q) = T_E \rtimes \operatorname{Aut}(E/\mathbb{F}_q, P_\infty),
\]
\[
|\operatorname{Aut}(E/\mathbb{F}_q)| = |T_E| \cdot |\operatorname{Aut}(E/\mathbb{F}_q, P_\infty)|.
\]
Moreover, the group law defined on $ \operatorname{Aut}(E/\mathbb{F}_q) $ is given by
\[
(\tau_P \alpha) * (\tau_Q \beta) = \tau_{P \oplus \alpha(Q)} \circ \alpha \beta.
\]
\end{theorem}

Thus, by applying Theorem~\ref{th:aut} together with the results of Section~\ref{sec:R-R-General_case}, we can efficiently construct quasi-cyclic elliptic codes with arbitrary order of quasi-cyclicity, provided that the curve possesses a point of the corresponding order.

\begin{example}

Let $\mathcal{E}$ be the elliptic curve defined over $\mathbb{F}_{59}$ by the equation $y^2 = x^3 + x + 1$. Consider the point $T = (1 : 11 : 1) \in \mathcal{E}(\mathbb{F}_{59})$ of order $\ell = 7$.  
The automorphism $\sigma = \tau_T$ (translation by $T$) belongs to $\operatorname{Aut}(\mathcal{E})$ and has order $\operatorname{ord}(\sigma) = 7$:
\[
(\tau_T)^j = \tau_{jT} \text{ and } \tau_T^j(P) = P \oplus jT.
\]

Take two points $P_1 = (0 : 1 : 1)$ and $P_2 = (0 : 58 : 1)$ in $\mathcal{E}(\mathbb{F}_{59})$.  
For each $i = 1,2$, define the orbit under $\sigma$ as
\[
\operatorname{Orb}_{\sigma}(P_i) = \{ P_i \oplus jT \mid j = 0,1,\dots,6 \},
\]
where $\oplus$ denotes the elliptic curve group law.  
These orbits are always disjoint and each consists of $7$ distinct rational points as for any $P,T \neq P_\infty$ and $j_1 \neq j_2$ with $j_i \leq \operatorname{ord}(T)$ holds  
\[
P \oplus j_1T = P \oplus j_2T \iff j_1 = j_2. 
\]

Set the divisor $D$ as the sum of these two orbits: $D = \sum_{i=1}^{2} \; \sum_{P \in \operatorname{Orb}_{\sigma}(P_i)} P$, so $\deg(D) = n = 14$. For the divisor $G$, choose a point $Q_1 = (11 : 24 : 1) \in \mathcal{E}(\mathbb{F}_{59})$ and take its $\sigma$-orbit:
\[
\operatorname{Orb}_\sigma(Q_1) = \{(11 : 24 : 1), (54 : 44 : 1), (49 : 17 : 1), (21 : 16 : 1), (26 : 27 : 1), (41 : 46 : 1), (15 : 21 : 1)\}
\]
\[
G = \sum_{Q \in \operatorname{Orb}_{\sigma}(Q_1)} Q.
\]

According to Theorem \ref{th:R-R_Arbitrary} and Algorithm \ref{alg:R-R_arbitrary}, one of the possible bases of the Riemann–Roch space has the form:
\[
\mathscr{L}_b =
\left\{
\begin{aligned}
1,\; &\frac{y+43x+23}{(x-11)(x-54)},
\frac{y+20x+40}{(x-11)(x-49)},
\frac{y+11x+21}{(x-11)(x-21)}, \\
&\frac{y+12x+10}{(x-11)(x-26)},
\frac{y+44x+12}{(x-11)(x-41)},
\frac{y+14x+47}{(x-11)(x-15)}
\end{aligned}
\right\}
\]

The evaluation map $\operatorname{ev}_D : \mathscr{L}(G) \to \mathbb{F}_{59}^{14}$ yields an $7$-quasi-cyclic AG code $\mathcal{C}_{\mathscr{L}}(D, G)$. This construction illustrates how Theorem~\ref{th:aut} together with the results of Section~\ref{sec:R-R-General_case} enables the efficient construction of quasi-cyclic elliptic codes with arbitrary quasi-cyclic order using an automorphism that does not fix the point at infinity.
\end{example}

\subsection{Goppa-like elliptic codes}\label{sec:Goppa-like}

In this section, we consider the construction of Goppa codes through the algebraic geometry approach, as well as its generalization to arbitrary subfield subcodes, which are called Goppa-like (see Definition \ref{def:Goppa-like-general}) algebraic geometry codes~\cite{KLN23}. It is well-known that Goppa codes are subfield subcodes of GRS codes, which in turn are algebraic geometry codes associated with the projective line of genus $g = 0$. Subsequently, in Lemma~\ref{lem:Goppa_AG_construction}, we present the form of divisors involved in the construction of Goppa codes. We note that this result is not new and can be found in~\cite{Stich90} as scattered lemmas and propositions. Here we present a revised concise proof for subsequent comparison with Goppa-like algebraic geometry codes.

\begin{lemma}[{\cite{Stich90}}]\label{lem:Goppa_AG_construction}
Let $D = P_1 + \ldots + P_n$ be a divisor whose support consists of $n$ pairwise distinct rational points of the projective line $\mathbf{P}^{1}$ over $\mathbb{F}_{q^m}$; $G=G_0-P_\infty$ be a divisor of the function field $\mathbb{F}_{q^m}(\mathbf{P}^{1})$ with disjoint support from $D$, where $G_0 = (g)_{0}$ and $g \in \mathbb{F}_{q}(\mathbf{P}^{1})$. Then the classical Goppa code can be represented in the following algebraic geometry form:
\begin{equation*}
    \Gamma(\mathbf{x}, g)=\mathcal{C}_{\mathscr{L}}(D,G_0 - P_{\infty})^{\perp} \cap \mathbb{F}_{q}^{n} = \mathcal{C}_{\mathscr{L}}(D, A-G_0) \cap \mathbb{F}_{q}^{n},
\end{equation*}
where $A = \operatorname{div}(h^\prime(z)) + (n-1)P_{\infty}$; $h(z) = \prod\limits_{x_i \in \mathbf{x}} (z-x_i)$.
\end{lemma}

\begin{proof}
Let $\mathbf{x} = \{ x(P_1),\ldots,x(P_n)\} = \{ x_1,\ldots,x_n \}$ and $g(z) \in \mathbb{F}_{q}(\mathbf{P}^{1})$, where $P_i = (x_i: 1) \in \mathbf{P}^1$ are pairwise distinct points; $\deg(g(z))=r$, $1 \leq r \leq n-1$; $g(x_i) \neq 0$ for all $x_i \in \mathbf{x}$. Consider $\text{GRS}_r(\mathbf{x},\mathbf{y})$ code with $\mathbf{y} = \left({g(x_i)}^{-1}\right)$ and generator matrix of the form:
\begin{equation*}
    V =
    \begin{pmatrix}
        g(x_1)^{-1} & g(x_2)^{-1} & \ldots & g(x_n)^{-1} \\
        x_1 g(x_1)^{-1} & x_2 g(x_2)^{-1} & \ldots & x_n g(x_n)^{-1} \\
        \vdots & \vdots & & \vdots \\
        x_1^{r-1} g(x_1)^{-1} & x_2^{r-1} g(x_2)^{-1} & \ldots & x_n^{r-1} g(x_n)^{-1}  
    \end{pmatrix}.
\end{equation*}

We will show that the following equality holds: 
\begin{equation}\label{Proof:eq}
    \text{GRS}_r(\mathbf{x},\mathbf{y}) = \mathcal{C}_{\mathscr{L}}(D,G) = \mathcal{C}_{\mathscr{L}}(D,G_0-P_\infty).
\end{equation}
To do this, it suffices to show that $z^j\cdot g(z)^{-1} \in \mathscr{L}(G_0-P_\infty)$ for all $j \in \{0,\ldots,r-1\}$. Consider the divisor of zeros and the divisor of poles of the function $z^j\cdot g(z)^{-1}$, then find its principal divisor:
\begin{equation*}
    \begin{split}
        &(z^j\cdot g(z)^{-1})_0 = jP_0 + rP_\infty - jP_\infty, \text{ where } P_0\text{ is the zero of the function } z;\\
        &(z^j\cdot g(z)^{-1})_\infty = G_0; \\
        &(z^j\cdot g(z)^{-1}) = j(P_0 - P_\infty) - (G_0 - rP_\infty) \geq -G_0 + P_\infty.
    \end{split}
\end{equation*}

Since $\dim(\mathscr{L}(G_0-P_\infty)) = r$, the basis of the Riemann–Roch space $\mathscr{L}(G_0-P_\infty)$ has the form $\{z^j\cdot g(z)^{-1}: j =0,\ldots,r-1\}$. Equality (\ref{Proof:eq}) is proved.

By Lemma 2.2.9 and Proposition 2.2.10 from~\cite{Stich90} we get:
\begin{equation}\label{Proof:eq_dual}
    \mathcal{C}_{\mathscr{L}}(D,G_0 - P_{\infty})^{\perp} = \mathcal{C}_{\mathscr{L}}(D, D - (G_0 - P_{\infty}) + \operatorname{div}(h^\prime(z)) - \operatorname{div}(h(z)) - 2P_\infty ),
\end{equation}
where $\operatorname{div}(h(z)) = (h(z))_0 - (h(z))_\infty = \sum\limits_{i=1}^{n} P_{i} - nP_\infty = D - nP_\infty$.
Thus, equality \eqref{Proof:eq_dual} becomes
\begin{equation*}
    \mathcal{C}_{\mathscr{L}}(D,G_0 - P_{\infty})^{\perp} =
    \mathcal{C}_{\mathscr{L}}(D, - G_0 + P_{\infty} + \operatorname{div}(h^\prime(z)) + nP_\infty - 2P_\infty ) = 
    \mathcal{C}_{\mathscr{L}}(D, A - G_0 ),
\end{equation*}
and therefore, the equality holds for the Goppa code:
\begin{equation*}
    \Gamma_q(\mathbf{x}, g)=\mathcal{C}_{\mathscr{L}}(D,G_0 - P_{\infty})^{\perp} \cap \mathbb{F}_{q}^{n} = \mathcal{C}_{\mathscr{L}}(D, A-G_0) \cap \mathbb{F}_{q}^{n}.
\end{equation*}
\end{proof}

As mentioned earlier, for most curves, the basis of the Riemann–Roch space has only been derived for divisors of the form $G = kP_\infty$ (or arbitrary one-point divisors). However, in the case of the projective line $\mathbf{P}^1$, computing the basis is trivial for any divisor, which allows for efficient construction of Goppa codes associated with divisors of the form $G = \operatorname{div}(g)_0 - P_{\infty}$, where $g \in \mathbb{F}_{q}(\mathbf{P}^{1})$.

Note that Goppa codes differ in their structure from alternant codes, as demonstrated in~\cite{MT23,BMT24}. Therefore, for cryptographic applications, it is best to consider codes whose structure is closest to that of Goppa codes, which after many years of research remain a secure option for use in code-based cryptosystems. In~\cite{KLN23}, the authors examine codes whose divisors match in form those used in the algebraic geometry construction of Goppa codes, and generalize the Schur product distinguisher~\cite{MT23} for alternant codes to the case of arbitrary subfield subcodes associated with $C_{a, b}$ curves.

\begin{definition}
Let $a, b$ be coprime positive integers. $C_{a, b}$ curve over $\mathbb{F}_{q^{m}}$ is a curve having an irreducible affine and non-singular plane model with equation

\begin{equation}\label{eq:Cab}
f_{a, b}(X, Y)=\alpha_{0 a} Y^{a}+\alpha_{b 0} X^{b}+\sum \alpha_{i j} X^{i} Y^{j}=0
\end{equation}
where $f_{a, b} \in \mathbb{F}_{q^{m}}[X, Y]$ with $\alpha_{0 a}, \alpha_{b 0} \neq 0$, and the sum is taken over all couples $(i, j) \in\{0, \ldots, b\} \times \{0, \ldots, a\}$ such that $a i+b j<a b$. The \textit{weighted degree} of the function is defined as follows: $\deg_{a,b}(f) = \deg_{a,b}(\operatorname{LT}(f))$, where $LT(f) = x^i y^j$ and $\deg_{a,b}(x^i y^j) = ai + bj$.
\end{definition}

Any curve $C_{a, b}$ defined as in Equation (\ref{eq:Cab}) has a unique point at infinity, denoted by $P_{\infty}$ and its genus is given by
\[
g_{a, b}=\frac{(a-1)(b-1)}{2}.
\]

Elliptic curves belong to the family of $C_{a,b}$ curves with parameters $a=2$ and $b=3$.

\begin{definition}[\cite{KLN23}]\label{def:Goppa-like-general}
Let $ G^\prime $ be an effective divisor on a smooth projective curve $ C_{a, b}/\mathbb{F}_{q^m} $, and $ g \in \mathbb{F}_{q^m}(C_{a, b}) \setminus \mathscr{L}(G^\prime) $. For the divisor $ D \in  \operatorname{Div}(C_{a, b}) $ with disjoint support from $ \operatorname{Supp}(G^\prime) \cup \operatorname{Supp}(\operatorname{div}(g)) $, define the AG code:
\begin{equation}\label{eq:Goppa-like_arbitrary}
    \mathcal{C} = \mathcal{C}_{\mathscr{L}}(D, G^\prime + \operatorname{div}(g)) = \left\{\operatorname{ev}_{D}(f g^{-1}) \mid f \in \mathscr{L}(G^\prime)\right\}.
\end{equation}
The associated Goppa-like AG code is:
\[
\Gamma(D, G^\prime, g) = \left.\mathcal{C}^\perp\right|_{\mathbb{F}_q}.
\]
\end{definition}

In~\cite{KLN23}, the authors specifically examine Goppa-like algebraic geometry codes with a fixed divisor $G = sP_\infty + \operatorname{div}(g)$, as the simplest case for constructing a basis of the Riemann–Roch space. The authors also demonstrate that the properties of elliptic codes with respect to the Schur product exactly coincide with those of alternant codes, indicating their structural similarity.

\begin{definition}[\cite{KLN23}]\label{def:Goppa-like_infinity}
  Let $s^{\prime}>s$ be two integers such that there exists a function $g \in \mathscr{L}\left(s^{\prime} P_{\infty}\right)$ with $\operatorname{deg}_{a, b}(g)=s^{\prime}$. Given a divisor $D \in \operatorname{Div}(C_{a, b})$ with disjoint support from $\operatorname{div}(g)$, we define the one-point Goppa-like AG code associated to $D, s$ and $g$ as
\begin{equation}\label{eq:Goppa-like_inf}
    \Gamma\left(D, s P_{\infty}, g\right)=\left.\mathcal{C}_{\mathscr{L}}\left(D,s P_{\infty}+\operatorname{div}(g)\right)^{\perp}\right|_{\mathbb{F}_{q}}
\end{equation}

\end{definition}

We now demonstrate that Definitions~\ref{def:Goppa-like-general} and~\ref{def:Goppa-like_infinity} generalizes the Goppa code construction from Lemma~\ref{lem:Goppa_AG_construction} to the case of elliptic curves, which are $C_{a,b}$ curves with $a = 2$, $b = 3$ and genus $g_{a,b} = 1$.

\begin{proposition}[Adapted from~\cite{KLN23}]\label{lem:Goppa-like-construction}
    Let $D = P_1 + \ldots + P_n \in \operatorname{div}(\mathcal{E})$ with support consists of $n$ pairwise distinct rational points of elliptic curve $\mathcal{E}/\FF_{q}$; $G=G_0-P_\infty \in \operatorname{div}(\mathcal{E})$, such that $\operatorname{supp}(D) \cap \operatorname{supp}(G) = \varnothing$, and $G_0 = (g)_{0}$ for some function $g \in \mathscr{L}\left((s+1) P_{\infty}\right) \setminus \mathscr{L}(sP_\infty)$. Then the  Goppa-like elliptic code $\left.\mathcal{C}_{\mathscr{L}}\left(D, sP_{\infty}+\operatorname{div}(g)\right)^{\perp}\right|_{\mathbb{F}_{q}} $ can be represented in the following algebraic geometry form:
\begin{equation*}
    \Gamma\left(D, s P_{\infty}, g\right)=\mathcal{C}_{\mathscr{L}}(D,G_0 - P_{\infty})^{\perp} \cap \mathbb{F}_{q}^{n},
\end{equation*}

\begin{proof}
    \[
\Gamma(D, s P_{\infty}, g) = \left.\mathcal{C}_{\mathscr{L}}\left(D, sP_{\infty}+\operatorname{div}(g)\right)^{\perp}\right|_{\mathbb{F}_{q}} = \left.\mathcal{C}_{\mathscr{L}}\left(D, sP_{\infty}+(g)_0 - (s+1)P_\infty\right)^{\perp}\right|_{\mathbb{F}_{q}} = \left.\mathcal{C}_{\mathscr{L}}\left(D, (g)_0 - P_\infty \right)^{\perp}\right|_{\mathbb{F}_{q}}.
\]
\end{proof}
\end{proposition}

This construction exactly coincides with the Goppa code formulation in Lemma~\ref{lem:Goppa_AG_construction}. However, the divisor $G^\prime$ is fixed as a multiple of the point at infinity to enable efficient construction of the considered codes, moreover, the authors impose an additional constraint on the function $g$ in examples, namely: $g = x^iy^j + g^\prime$ for $g^\prime \in \mathscr{L}(sP_\infty)$ and some $i, j$, chosen to satisfy the condition on degree, i.e. $ai + bj = s+1$. Thus, even in this special case, we encounter significant limitations in selecting appropriate functions. This makes the derivation of bases for Riemann–Roch spaces associated with arbitrary divisors of Goppa-like elliptic codes an important research problem. Such a result could potentially enable efficient construction of codes for which the distinguisher from~\cite{KLN23} would be completely inapplicable.

If we abstract away from the definition of Goppa-like AG codes and focus on the core concept that the finite divisor must take the form $G = G_0 - P_\infty$ to ensure that functions in the Riemann–Roch space are rational rather than polynomial, we can in practice construct codes using more general divisors by applying Theorem \ref{th:R-R_Arbitrary}. In this case we can choose $G_0 = k_1P_1 + \dots k_{\tau}P_{\tau}$ with $\deg(G_0) = s+1$ and without any restrictions, except $k_i \in \NN_{>0}$.

\begin{remark}
The basis functions introduced in Theorem~\ref{th:R-R_Arbitrary} remain valid for divisors of the form $ G = G_0 - P_\infty $, where $ G_0 = \sum_i k_iP_i $ is an effective divisor and $P_i^\prime = -P_i$ denotes the negative of $P_i$ in the group law:
\[
\operatorname{div}\left(\frac{Y + A_{i,j}(X)}{(X - \alpha_i)^j}\right) \geq 2jP_\infty - jP_i - \cancel{jP_{i}^\prime} + \cancel{jP_{i}^\prime} - 2(j-1)P_\infty \geq -jP_i + P_\infty \stackrel{j \leq k_i}{\geq} -G,
\]
\[
\operatorname{div}\left(\frac{Y + B_i(X)}{(X - \alpha_i)(X - \alpha_{i+1})}\right) \geq 4P_\infty - \sum_{j = i}^{i+1} \left(P_{j} + P_j^\prime\right) + \sum_{j = i}^{i+1} P_j^\prime +  sP_{\alpha, \beta^\prime} - 3P_\infty \geq -P_i - P_{i+1} + P_\infty \geq -G,
\]
\[
\operatorname{div}\left(\frac{1}{X - \alpha_i}\right) \geq 2P_\infty -P_i - P_{i+1} \geq -G \quad (\text{if } \alpha_i = \alpha_{i+1}).
\]
\end{remark}

If we consider general Goppa-like elliptic code, it is obvious that we can also use the results from Section \ref{sec:R-R-General_case} to build Riemann–Roch space bases, associated with divisors $G = G^\prime + \operatorname{div}(g)$ as functions in this case are of the form: $f = f^\prime/g$, where $f^\prime \in \mathscr{L}(G^\prime)$ and $G^\prime$ is effective. The basis functions of $\mathscr{L}(G^\prime + \operatorname{div}(g))$ will take the form:
\[
f_{i,s}(X,Y) = \frac{Y + A_{i,s}(X)}{(X - \alpha_i)^s \cdot g}, \quad
g_i(X,Y) = \frac{Y + B_i(X)}{(X - \alpha_i)(X - \alpha_{i+1})\cdot g} \quad (\text{see Theorem \ref{th:R-R_Arbitrary}})
\]

To simplify the construction, following the analogy with one-point codes, we can consider divisors satisfying $\mathscr{L}(G') \subset \mathscr{L}(\operatorname{div}(g))$ with $g \not\in  \mathscr{L}(G')$. This case is completely analogous to the previously considered situation for one-point Goppa-like elliptic codes, and therefore we will omit it in the present work.

Now we will demonstrate the possibility of effective construction QC-Goppa-like AG codes by deriving necessary constraints on functions and divisors that guarantee the quasi-cyclicity property of the associated codes.

\begin{proposition}\label{prop:QC-Goppa-like}
Let $D, G^\prime$ be divisors of $\mathcal{E}/\FF_{q}$ of the form $D = \textstyle\sum\limits_{i=1}^{n/\ell} \sum\limits_{P \in \operatorname{Orb}_{\sigma}(P_{i})} P$, and $G^\prime = \sum_{z=1}^{s} t_{z} \sum_{Q \in \operatorname{Orb}_{\sigma}(Q_{z})} Q$, where $s \in \mathbb{N}$, $\sigma \in \operatorname{Aut}(\mathcal{E}/\FF_q, P_\infty)$ and $Q_z \in \mathcal{E}(\FF_{p^m})\backslash (\{P_\infty, P_{0, \beta_1}, P_{0, \beta_2}, P_{\alpha_1, 0}, P_{\alpha_2, 0}, P_{\alpha_3, 0}\} \cup \mathcal{E}[3])$. If function $g \not\in \mathscr{L}(G^\prime)$ and $g$ is of the form: 
\[
g = \textstyle\prod_{z = 1}^{s} \left[\textstyle\prod_{\rho = 1}^{\ell/2}\left(X-x(\mathcal{Q}_{z,\rho})\right)\right]^{t_z^\star}, \text{ where }
\begin{cases}
     t_z^\star = t_z + 1 & \text{if } \forall \mathcal{Q}_{z,\rho} \in \operatorname{supp}(G^\prime)\\
    t_z^\star > 0 & \text{if } \exists \mathcal{Q}_{z,\rho} \not\in \operatorname{supp}(G^\prime)
\end{cases},
\]
then the Goppa-like code $\mathcal{C}_{\mathscr{L}}(D, G^\prime + \operatorname{div}(g))$ is quasi-cyclic

\begin{proof}
The reasoning is the same as in the general case of Goppa-like elliptic codes, and the result follows directly from Theorem~\ref{th:R-R-QC} and Definition~\ref{def:Goppa-like-general}.
\end{proof}
\end{proposition}

\begin{remark}
Using Theorem~\ref{th:aut} and the result of Section~\ref{sec:R-R-General_case}, it is possible to construct both Goppa-like codes and quasi-cyclic codes associated with arbitrary divisors, whose Riemann–Roch space bases consist of rational functions.
\end{remark}

\paragraph{Cryptographic application of Goppa-like algebraic geometry codes}

In~\cite{KLN23} the dimension of the square of the dual Goppa-like AG codes is considered. Authors prove the upper bounds on dimensions of squared dual codes which lead to the effective (polynomial) distinguisher. Let $k = \dim_{\FF_{q^m}}\left(\mathcal{C}_{\mathscr{L}}\left(D,G^\prime+\operatorname{div}(g)\right)\right)$, $s = \deg(G^\prime) > g_{a, b}$, where $G^\prime$ is effective divisor. Then dimensions of the codes admit the following bounds:
\begin{equation}\label{eq:Goppa-like-dim}
\begin{gathered}
\dim(\Gamma(D, G^\prime, g)^\perp)^{\star 2} \leq  \binom{mk + 1}{2}  - \frac{m}{2} \left( k(k - 1)(2e + 1) - 2s \left( \frac{q^{e+1} - 1}{q - 1} \right) \right)\\
\operatorname{dim}(\Gamma\left(D, s_\infty P_{\infty}, g^*\right)^{\perp})^{\star 2} \leq   \binom{mk + 1}{2} - \frac{m}{2} (k^2 (2e^* + 1) + k - 2s^* (q^{e^*} - q^{e^* - 1} + 1)) 
\end{gathered}
\end{equation}
where $s^* = \deg_{a,b}(g^*)$, $s_\infty \geq \left(s^{*}-s_\infty\right) q+2 g_{a, b}-1$; $e=\min \left(\left\lfloor\frac{m}{2}\right\rfloor,\left\lceil\log _{q}\left(\frac{k^{2}}{s}\right)\right\rceil\right)$; $e^{*}=\min \left(\left\lfloor\frac{m}{2}\right\rfloor,\left\lceil\log _{q}\left(\frac{k^{2}}{s^{*}(q-1)^{2}}\right)\right\rceil+1\right)$.

Thus, the code is distinguishable from a random one if the right hand-side of (\ref{eq:Goppa-like-dim}) is smaller than the length $n$ of the code. This holds because for a random code the dimension of its squared dual code is likely to be equal $\binom{mk+1}{2}$~\cite{MT23}.

In~\cite{KLN23}, the authors compute the maximal distinguishable value of $s_\infty$ for a specific family of elliptic codes whose parameters are suitable for cryptographic applications. They consider the case of the dual code $\Gamma\left(D, s_\infty P_\infty, g\right)^{\perp}$ with $n = |\mathcal{E}(\mathbb{F}_{q^m})| - 2$ and $|\operatorname{supp}(\operatorname{div}(g))| = 2$, fixing $s^* = s_\infty + 1$. The resulting values are presented in Table~\ref{table:Distinguisher}. It is worth noting that these findings coincide with known distinguisher results for random alternant and Goppa codes of high rate~\cite{MT23, FOPT11}.

\begin{table}[ht!]
\centering
\begin{tabular}{ccccc}
\hline$q$ & $m$ & $n$ & Largest distinguishable $s_\infty$ & Corresponding rate \\
\hline 2 & 12 & 4218 & 14 & 0,963 \\
\hline 2 & 13 & 6688 & 18 & 0,982 \\
\hline 3 & 7 & 2186 & 15 & 0,962 \\
\hline 3 & 8 & 6393 & 24 & 0,977 \\
\hline 7 & 4 & 2395 & 27 & 0,957 \\
\hline 7 & 5 & 4650 & 26 & 0,971 \\
\hline 7 & 5 & 8192 & 37 & 0,979 \\
\hline
\end{tabular}
\caption{Largest distinguishable one-point Goppa-like AG code in elliptic case from~\cite{KLN23}}\label{table:Distinguisher}
\end{table}

We expect that potentially the distinguisher will succeed with smaller values of $s$ for multipoint divisors than for the one-point case, however we leave investigation to future research. The result of Proposition~\ref{prop:QC-Goppa-like} is particularly interesting from a cryptographic perspective, as it establishes a basis for the Riemann–Roch space of divisors in QC-Goppa-like elliptic codes. The integration of this code family could potentially enhance the scheme's security compared to common QC codes.

\subsection{Quasi-cyclic MDS and iso-dual elliptic codes}\label{sec:mds_iso-dual}

Linear codes that combine optimal distance properties with strong algebraic symmetry form a fundamental object of study in coding theory. Special emphasis is placed on maximum distance separable (MDS) codes \cite{DY23,LWZ15,M92}, which satisfy the Singleton bound $d = n - k + 1$ and consequently guarantee optimal error-correction. At the same time, substantial research has addressed isometry-dual codes (iso-dual) \cite{ZZ25,CPQT24,ST88,BDH20,BCQ22,BCQ23,KL17}, i.e., codes equivalent to their dual $\cC^\perp$. This notion generalizes more traditional families: self-dual codes ($\cC = \cC^\perp$) and linear complementary dual (LCD) codes ($\cC \cap \cC^\perp = {0}$).

We denote by $\oplus$ the addition operation in the group of points of an elliptic curve (or its function field). The following lemmas describe known restrictions that must be imposed on divisors to achieve the MDS property.

\begin{lemma}[{\cite{CH24}}]\label{lemma:2.5}
Suppose that $ G $ can be represented by $ G = Q_1 + \ldots + Q_k $ where $ Q_1, \ldots, Q_k $ are $ k $ points and distinct from $ P_i $ for all $ i $. Then the algebraic geometry code $ \mathcal{C}_{\mathscr{L}}(D, G) $ is MDS if and only if  
\[
P_{i_1} \oplus \ldots \oplus P_{i_k} \neq Q_1 \oplus \ldots \oplus Q_k
\]
for any $ k $ distinct points $ P_{i_1}, \ldots, P_{i_k} $ in $ \operatorname{supp}(D) $.
\end{lemma}

\begin{lemma}[{\cite[Lemma 2.6]{ZZ25}}]\label{lemma:2.6}
Suppose that $Q_1, \ldots, Q_s$ are distinct points such that the order of $Q_j$ is prime to the order of $P_i$ for all $1 \leq i \leq n$ and $1 \leq j \leq s$. Denote by
\[
\mathcal{Q}_j = \{Q_j \oplus P_i : 1 \leq i \leq n\}.
\]
Fix the divisors $D = \sum_{j=1}^{s} \sum_{P \in \mathcal{Q}_j} P$ and $G = k P_\infty$. If
\[
[k_1]Q_1 \oplus \ldots \oplus [k_s]Q_s \neq  P_\infty
\]
for all non-negative integers $k_1,\ldots,k_s$ with
\[
k_1 + \ldots + k_s = k,
\]
then the algebraic geometry code $\cC_\mathscr{L}(D, G)$ is an $[ns, k]$ MDS code.
\end{lemma}

\begin{corollary}[{\cite[Corollary 2.1]{ZZ25}}]\label{cor:2.1}
With the notations in Lemma~\ref{lemma:2.6}, suppose that there exist $t$ points $Q'_1, \ldots, Q'_t$ such that $Q'_j \notin \operatorname{supp}(D)$ and the order $q_j$ of $Q'_j$ is prime to $P_i$ for all $1 \leq i \leq n$ and $1 \leq j \leq t$. Let
\[
\tilde{G} = m_1 Q'_1 + \ldots + m_t Q'_t + (k - m_1 - \ldots - m_t) P_\infty.
\]
If
\[
[k_1]Q_1 \oplus \ldots \oplus [k_s]Q_s \oplus [q_1 - m_1]Q'_1 \oplus \ldots \oplus [q_t - m_t]Q'_t \neq  P_\infty
\]
for all
\[
k_1 + \ldots + k_s = k,
\]
then the code $\mathcal{C}_{\mathscr{L}}(D, \tilde{G})$ is an $[ns, k]$ MDS code.
\end{corollary}

\subsubsection{Quasi-cyclic MDS elliptic codes}

The literature contains extensive studies of MDS elliptic codes. In particular, \cite{MUNUERA93, CH24, ZZ25} establish restrictions on divisors for constructing codes that possess this property. Although the works \cite{CH24,ZZ25} derive restrictions on arbitrary divisors $G$ for achieving the MDS property, special attention is given to cases where the bases of the Riemann–Roch space can be computed trivially. In particular, examples are provided for divisors of the form $G = kP_\infty$ and $G = sQ + (k-s)P_\infty$, where $Q$ is a $2$-torsion point. In the latter case, the basis of the Riemann–Roch space consists of the basis functions from $\mathscr{L}((k-s)P_\infty)$ together with additional functions $g_i = \frac{1}{(X - x(Q))^i}$, whose principal divisor is $\operatorname{div}(g_i) = 2iP_\infty - 2iQ$. Our results in Section \ref{sec:R-R-General_case} allow us to construct codes associated with arbitrary divisors, therefore making it possible to efficiently use constructions of MDS codes associated with divisors having a large number of points. Next, we present the construction of a quasi-cyclic MDS elliptic code with more general divisors, obtained via an action of automorphism $\sigma \in \operatorname{Aut}(\mathcal{E}/\mathbb{F}_q, P_\infty)$ that fixes the point at infinity.

\begin{theorem}\label{thm:QC-MDS-elliptic-general}
Let $\mathcal{E}/\mathbb{F}_q$ be an elliptic curve and let $\sigma \in \operatorname{Aut}(\mathcal{E}/\mathbb{F}_q, P_\infty)$ be an automorphism of order $\ell \geq 2$.
Let $Q_1,\dots,Q_s$ ($s\geq 1$) be pairwise distinct points fixed by $\sigma$ and different from $ P_\infty$.
Let $P_1,\dots,P_r$ be points whose $\sigma$-orbits are disjoint and have size $\ell$, and assume that the order of each $P_i$ is coprime to the order of every $Q_j$.
Define the divisor
\[
D = \sum_{j=1}^{s} \sum_{i=1}^{r} \sum_{k=0}^{\ell-1} \bigl(Q_j \oplus \sigma^{k}(P_i)\bigr),
\]
so that $\deg(D) = s r \ell$.
Let $Q'$ be a point whose orbit under $\sigma$ has size $\ell$ and denote its orbit by
\[
\operatorname{Orb}_\sigma(Q') = \{Q'_0 = Q',\; Q'_1 = \sigma(Q'),\; \dots,\; Q'_{\ell-1} = \sigma^{\ell-1}(Q')\}.
\]
Assume that the sum of the points in this orbit equals $ P_\infty$, i.e.
\[
\bigoplus_{j=0}^{\ell-1} Q'_j =  P_\infty,
\]
that the order $q'$ of $Q'$ is coprime to the orders of all $P_i$, and that $\operatorname{supp}(D) \cap \operatorname{Orb}_\sigma(Q') = \varnothing$.
Choose an integer $m$ with $0 \leq m \leq \lfloor k/\ell \rfloor$ and define
\[
G = m \sum_{j=0}^{\ell-1} Q'_j \;+\; (k - m\ell) P_\infty,
\]
so that $\deg(G) = k < sr\ell$.
If for every $s$-tuple of non‑negative integers $(k_1,\dots,k_s)$ with $k_1+\ldots +k_s = k$ we have
\begin{equation}\label{eq:condition-QC-MDS-general}
[k_1]Q_1 \oplus \ldots \oplus [k_s]Q_s \;\neq\;  P_\infty,
\end{equation}
then the algebraic geometry code $\cC_\mathscr{L}(D, G)$ is an $\ell$-quasi‑cyclic $[sr\ell,\;k]$ MDS code.
\end{theorem}

\begin{proof}
The divisor $D$ is a union of $sr$ disjoint $\sigma$-orbits of size $\ell$, each orbit being of the form $\{Q_j \oplus \sigma^k(P_i) : k = 0,\dots,\ell-1\}$.  
The divisor $G$ is $\sigma$-invariant because it is a linear combination of the full orbit $\operatorname{Orb}_\sigma(Q')$ (which is permuted by $\sigma$) and the fixed point $ P_\infty$.  
Hence, by Definition~\ref{def:QC-Elliptic}, $\cC_\mathscr{L}(D,G)$ is $\ell$-quasi‑cyclic.

To prove the MDS property we apply Corollary~\ref{cor:2.1}.  
In the notation of the corollary we set:
\begin{itemize}
    \item $\mathcal{Q}_j = \{Q_j \oplus \sigma^k(P_i) : 1\leq i\leq r,\;0\leq k\leq \ell-1\}$ for $j=1,\dots,s$;
    \item $Q'_1,\dots,Q'_\ell$ as the points of $\operatorname{Orb}_\sigma(Q')$ (so $t=\ell$);
    \item $m_1 = \ldots = m_\ell = m$.
\end{itemize}
All required coprimality conditions are satisfied as the orders of $Q_j$ and $P_i$ are coprime, and the order $q'$ of $Q'$ is coprime to the orders of $P_i$.

Since the sum of the orbit of $Q'$ is $ P_\infty$, we have
\[
\bigoplus_{j=0}^{\ell-1} [q'-m]Q'_j = [q'-m]\Bigl(\bigoplus_{j=0}^{\ell-1} Q'_j\Bigr) = [q'-m] P_\infty =  P_\infty.
\]
Therefore the condition required in Corollary~\ref{cor:2.1} reduces to
\[
[k_1]Q_1 \oplus \ldots \oplus [k_s]Q_s \;\neq\;  P_\infty
\]
for all $k_1+\ldots+k_s = k$, which is exactly condition \eqref{eq:condition-QC-MDS-general}.  
Consequently, $\cC_\mathscr{L}(D,G)$ is an $[sr\ell,k]$ MDS code.
\end{proof}

\begin{corollary}\label{cor:QC-MDS-elliptic-single}
Consider the condition of Theorem~\ref{thm:QC-MDS-elliptic-general} with $s = 1$.  
Let $Q = Q_1$ be a point fixed by $\sigma$, and keep the same notation for $P_i$, $Q'$, $m$, $k$.  
If
\[
[k]Q \neq  P_\infty,
\]
then the code $\cC_\mathscr{L}(D,G)$ is an $\ell$-quasi‑cyclic $[r\ell,k]$ MDS code.
\end{corollary}

\begin{proof}
This is the special case $s=1$ of Theorem~\ref{thm:QC-MDS-elliptic-general}.  
Condition \eqref{eq:condition-QC-MDS-general} becomes simply $[k]Q \neq  P_\infty$, and the length simplifies to $1 \cdot r \ell = r\ell$.
\end{proof}

\begin{remark}
The condition $[k]Q \neq  P_\infty$ is easy to satisfy in practice:
\begin{itemize}
    \item If $Q$ has order $2$, then $[k]Q \neq  P_\infty$ means $k$ must be odd.
    \item If $Q$ has order $3$, then $[k]Q \neq  P_\infty$ means $k$ cannot be a multiple of $3$.
\end{itemize}
This provides a simple criterion for constructing $\ell$-quasi-cyclic MDS codes from elliptic curves.
\end{remark}


\begin{example}\label{ex:QC-MDS-l3}
Consider an elliptic curve $\mathcal{E}$ over $\mathbb{F}_q$ with $p = \operatorname{char}(\mathbb{F}_q) > 3$ and an automorphism $\sigma$ of order $\ell=3$. By Table~\ref{tab:invariant_points_orbits}, the fixed points of $\sigma$ are exactly the $3$-torsion points. Let $Q$ be a non‑zero $3$-torsion point, so $Q$ has order $3$ and is fixed by $\sigma$. Choose $r$ points $P_1,\dots ,P_r$ and point $Q'$ whose $\sigma$-orbits have size $3$ and whose orders are coprime to $3$. For any dimension $k$ that is not a multiple of $3$, condition \eqref{eq:condition-QC-MDS-general} holds because $[k]Q = Q$ if $k \equiv 1 \pmod{3}$ and $[k]Q = 2Q$ if $k \equiv 2 \pmod{3}$, both different from $ P_\infty$. Then, for any $m$ with $0 \leq m \leq \lfloor k/3 \rfloor$, the code $\cC_\mathscr{L}(D, G)$ with $D$ and $G$ defined as in Corollary~\ref{cor:QC-MDS-elliptic-single} is a $[3r, k]$ MDS $3$-quasi-cyclic code.
\end{example}

\subsubsection{Quasi-cyclic iso-dual codes}

In the work \cite{ZZ25}, the authors examine several constructions of iso-dual elliptic MDS codes. Two linear codes $ \mathcal{C}_1 $ and $ \mathcal{C}_2 $ are called equivalent if there exists a nonzero rescaling vector $ v \in (\mathbb{F}_q^*)^n $ and a coordinate permutation such that
\[
\mathcal{C}_2 = \{(c_1, \dots, c_n) \mid c_i = v_i x_i, \; (x_1, \dots, x_n) \in \operatorname{Per}(\mathcal{C}_1)\},
\]
where $\operatorname{Per}(\mathcal{C}_1)$ is the code obtained from $\mathcal{C}_1$ by a coordinate permutation. A code $\mathcal{C}$ is \textit{isometry-dual} (iso-dual) if it is equivalent to its dual code $\mathcal{C}^\perp$.

We further show that the codes considered in the examples of \cite{ZZ25} possess the property of quasi-cyclicity, and also present more general methods for constructing quasi-cyclic iso-dual elliptic codes with MDS property.

\begin{proposition}[Adapted from \cite{ZZ25}]\label{prop:qc-construction1}
Let $q$ be a power of $2$, and let $\mathcal{E}/\mathbb{F}_q$ be an elliptic curve with a rational point $Q_1 \in \mathcal{E}[2]$. Let $k$ be an even integer, and let the points $\{P_{i}$, $P'_{i} = -P_{i}\}_{i=1,\dots,k}$  be non-invariant points under the action of involution. Define the divisors  
\[
D = \sum_{i=1}^k \big( (Q_1 \oplus P_{i}) + (Q_1 \oplus P'_{i}) \big), \qquad G = (k-1) P_\infty + Q_1.
\]  
Then the code $\cC_\mathscr{L}(D, G)$ is a $[2k, k]$ MDS iso-dual code that is 2-quasi-cyclic with respect to the automorphism $\sigma(P) = -P$.
\end{proposition}

\begin{proof}
 The code $\cC_\mathscr{L}(D, G)$ is an MDS iso-dual by \cite[Theorem 4.1]{ZZ25} . To see the quasi-cyclic structure, note that for each $i$,  
\[
\sigma(Q_1 \oplus P_{i}) = -Q_1 \oplus (-P_{i}) = Q_1 \oplus P'_{i},
\]  
since $Q_1$ has order $2$. Thus the set $\{Q_1 \oplus P_{i}, Q_1 \oplus P'_{i}\}$ is exactly the orbit of $Q_1 \oplus P_{i}$ under $\sigma$. Hence $D$ is a sum of $k$ disjoint orbits of size $2$. Moreover, both $ P_\infty$ and $Q_1$ are fixed by $\sigma$, so the divisor $G$ is $\sigma$-invariant. Therefore, by Definition \ref{def:QC-Elliptic}, the code $\cC_\mathscr{L}(D, G)$ is $2$-quasi-cyclic.
\end{proof}

\begin{proposition}[Adapted from \cite{ZZ25}]\label{prop:qc-construction2}
Let $q$ be odd, and let $\mathcal{E}/\mathbb{F}_q$ be an elliptic curve with $\mathcal{E}[2] \subseteq \mathcal{E}(\mathbb{F}_q)$. Let $k$ be an even integer, and let the points $\{P_{i}$, $P'_{i} = -P_{i}\}_{i=1,\dots,k}$ be defined as in Proposition \ref{prop:qc-construction1}. Define the divisors  
\[
D = \sum_{i=1}^{k/2} \Big( (Q_1 \oplus P_{i}) + (Q_1 \oplus P'_{i}) + (Q_2 \oplus P_{i}) + (Q_2 \oplus P'_{i}) \Big), \qquad G = (k-1) P_\infty + Q_1,
\]  
where $Q_1, Q_2 \in \mathcal{E}[2]$ are distinct. Then the code $\cC_\mathscr{L}(D, G)$ is a $[2k, k]$ MDS iso-dual code that is 2-quasi-cyclic with respect to the automorphism $\sigma(P) = -P$.
\end{proposition}

\begin{proof}
By \cite[Theorem 5.1]{ZZ25}, the code $\cC_\mathscr{L}(D, G)$ is MDS iso-dual. For each $i$ and $j=1,2$, we have  
\[
\sigma(Q_j \oplus P_{i}) = -Q_j \oplus (-P_{i}) = Q_j \oplus P'_{i},
\]  
since $2$-torsion points satisfy $-Q_j = Q_j$. Thus each pair $\{Q_j \oplus P_{i}, Q_j \oplus P'_{i}\}$ is an orbit under $\sigma$. Consequently, $D$ is a union of $k$ disjoint orbits of size $2$. The divisor $G$ is $\sigma$-invariant because $ P_\infty$ and $Q_1$ are fixed by $\sigma$. Hence the code is $2$-quasi-cyclic by Definition \ref{def:QC-Elliptic}.
\end{proof}

As noted earlier, the bases of the Riemann–Roch space under the conditions of Proposition~\ref{prop:qc-construction1} and Proposition~\ref{prop:qc-construction2} are trivial and are derived in~\cite{ZZ25}. Using the following proposition, we will consider more general approaches to constructing quasi-cyclic iso-dual codes with order of quasi-cyclicity $\ell > 2$, possessing the MDS property.

\begin{proposition}[{\cite[Proposition 2.6]{CPQT24}}]\label{prop:iso-dual}
 Let $ \cC_{\mathscr{L}}(D, G) $ be an AG-code of length $ n $ over a function field $ \mathcal{F} $ of genus $ g $. Code $ \cC_{\mathscr{L}}(D, G) $ is iso-dual if there exists a canonical divisor $ W $ such that $ W \sim 2G - D $. More precisely,
    \[
    \cC_{\mathscr{L}}(D, G)^{\perp} = x \cdot \cC_{\mathscr{L}}(D, G)
    \]
    where $ x = (\operatorname{res}_{P_1}(\omega), \ldots, \operatorname{res}_{P_n}(\omega)) $ with $ \omega $ a Weil differential such that $ W \sim (\omega) $.
 
\end{proposition}

\begin{theorem}[{\cite[Theorem 11.2]{Wash08}}]\label{th:pricipal}
Let $ D = \sum_{i=1}^n a_i P_i $ be a divisor with $ \deg(D) = 0 $, where $ P_i $ are all rational places. Then $ D $ is a principal divisor, i.e., there exists a function $ f $ such that
\[
(f) = D
\]
if and only if $ \operatorname{sum}(D) =  P_\infty $.
\end{theorem}

\begin{theorem}\label{th:qc-isodual-translation}
Let $\mathcal{E}/\mathbb{F}_q$ be an elliptic curve and let $T\in\mathcal{E}(\mathbb{F}_q)$ be a point of odd order $\ell\geq 3$.
Define the translation automorphism $\sigma(P)=P\oplus T$ with order $\ell$.
Assume there exist an odd integer $r\geq 1$ and points $P_1,\dots,P_r\in\mathcal{E}(\mathbb{F}_q)$ satisfying
\[
\bigoplus_{i=1}^r P_i =  P_\infty,
\]
each $P_i$ has order coprime to $\ell$, and the $2r\ell$ points
$A\oplus P_i\oplus jT$ and $(-A)\oplus P_i\oplus jT$ ($1\leq i\leq r$, $0\leq j\leq\ell-1$) are pairwise distinct and disjoint from the orbit $\{jT:0\leq j\leq\ell-1\}$.
Assume further that there exists a point $A\in\mathcal{E}(\mathbb{F}_q)$ of order $d$ with $\gcd(d,\ell)=1$, $\gcd(d,\operatorname{ord}(P_i))=1$ for every $i$, and $d > r\ell$.
Define divisors
\[
D = \sum_{i=1}^r\sum_{j=0}^{\ell-1}\Bigl((A\oplus P_i\oplus jT)+((-A)\oplus P_i\oplus jT)\Bigr),\qquad \deg(D) = 2r\ell,
\]
\[
G = r P_\infty + r\sum_{j=1}^{\ell-1}(jT),\qquad \deg(G) = r\ell.
\]
Then the algebraic geometry code $\mathcal{C}_{\mathscr{L}}(D,G)$ is an $[n,k,n-k+1]$ MDS code with $n=2r\ell$, $k=r\ell$, which is $\ell$-quasi-cyclic and iso-dual.
In particular, $\mathcal{C}_{\mathscr{L}}(D,G)^{\perp} = \mathbf{v}\mathcal{C}_{\mathscr{L}}(D,G)$ for some vector $\mathbf{v}\in(\mathbb{F}_q^{\ast})^n$.
\end{theorem}

\begin{proof}
Divisors $D$ and $G$ are unions of orbits formed by the translation map: $D$ consists of the orbits of $A\oplus P_i$ and $-A\oplus P_i$ (each of size $\ell$, since T has order $\ell$), and $G$ is the orbit of $ P_\infty$ taken with multiplicity $r$.
Hence $\sigma(D)=D$ and $\sigma(G)=G$.
By Definition~\ref{def:QC-Elliptic}, $\mathcal{C}_{\mathscr{L}}(D,G)$ is $\ell$-quasi-cyclic.

By the conditions of the theorem, $\ell$ is odd, hence $\bigoplus_{j=0}^{\ell-1}jT =  P_\infty$ and for any point $Q$ the sum of its orbit is $\ell Q$. Therefore
\[
\bigoplus\operatorname{supp}(D)= \bigoplus_{i=1}^r\bigl(\ell(A\oplus P_i)+\ell((-A)\oplus P_i)\bigr)
= 2\ell\bigoplus_{i=1}^r P_i =  P_\infty,
\]
and
\[
\bigoplus\operatorname{supp}(G)= r\ell P_\infty=  P_\infty,\qquad 
\bigoplus\operatorname{supp}(2G)=2 P_\infty= P_\infty.
\]
Since $\bigoplus(2G-D)= P_\infty$, Theorem \ref{th:pricipal} implies that the divisor $2G-D$ is principal of degree $\deg(2G-D)= 0$. By Proposition \ref{prop:iso-dual}, the code $\mathcal{C}_{\mathscr{L}}(D,G)$ is iso-dual, consequently, there exists a Weil differential $\eta$ with $(\eta)\sim 2G-D$ such that
$\mathcal{C}_{\mathscr{L}}(D,G)^{\perp}=\mathcal{C}_{\mathscr{L}}(D,G,\mathbf{v})$, where $\mathbf{v}=(\operatorname{res}_{P_i}(\eta))_i$.

By Lemma~\ref{lemma:2.5}, the code $\mathcal{C}_{\mathscr{L}}(D,G)$ is MDS if and only if for every choice of $k$ distinct points
\[
P'_1,\dots,P'_k\in \operatorname{supp}(D),
\]
we have
\[
P'_1\oplus\ldots\oplus P'_k \ne  P_\infty.
\]

Let $P_1',\dots,P_k'$ be any $k=r\ell$ distinct points from $\operatorname{supp}(D)$.  
Each point in $\operatorname{supp}(D)$ is of the form
\[
A\oplus P_i\oplus jT
\quad \text{or} \quad
-A\oplus P_i\oplus jT,
\]
for some $1\leq i\leq r$ and $0\leq j\leq \ell-1$.

Let $k_1$ denote the number of selected points of the form $A\oplus P_i\oplus jT$, and let $k_2$ denote the number of selected points of the form $-A\oplus P_i\oplus jT$. Then
\[
k_1+k_2=k=r\ell.
\]

Summing the selected points gives
\[
P'_1\oplus\ldots\oplus P'_k
=
[k_1]A \oplus [k_2](-A)
\oplus
\bigoplus_{t=1}^{k}(P_{i_t}\oplus j_t T).
\]

Hence
\[
P'_1\oplus\ldots\oplus P'_k
=
[k_1-k_2]A
\oplus
\bigoplus_{t=1}^{k}(P_{i_t}\oplus j_t T).
\]

Since $k=r\ell$ is odd, we have $k_1\ne k_2$, and therefore
\[
0<|k_1-k_2|\leq k.
\]

Because $\operatorname{ord}(A)\geq k$, it follows that
\[
[k_1-k_2]A\ne  P_\infty.
\]

Now observe that the subgroup generated by the points $P_i$ and $T$ has order dividing $\operatorname{lcm}(\{\operatorname{ord}(P_i)\},\ell)$. Because $\gcd(d,\ell)=1$ and $\gcd(d,\operatorname{ord}(P_i))=1$ for every $i$, these two subgroups intersect only in $ P_\infty$. Consequently, the element $[k_2-k_1]A$ cannot be cancelled by any element of the form $\bigoplus_{t=1}^{k}(P_{i_t}\oplus j_t T)$. Thus, for every choice of $k$ distinct points in $\operatorname{supp}(D)$,
\[
P'_1\oplus\ldots\oplus P'_k \ne  P_\infty.
\]

By Lemma~\ref{lemma:2.5}, the code $\mathcal{C}_{\mathscr{L}}(D,G)$ is MDS.
\end{proof}

\begin{theorem}\label{th:iso-dual_order_2}
Let $\mathcal{E}/\mathbb{F}_q$ be an elliptic curve and let $T\in\mathcal{E}(\mathbb{F}_q)$ be a point of odd order $\ell\geq 3$.
Let $Q_1\in\mathcal{E}(\mathbb{F}_q)$ be a point of order $2$. Assume there exist an integer $r=2s$ with $s$ odd, and points $P_1,\dots,P_r\in\mathcal{E}(\mathbb{F}_q)$ such that:
\begin{enumerate}
    \item $\displaystyle\bigoplus_{i=1}^r P_i =  P_\infty$;
    \item each $P_i$ has order coprime to $\ell$ and $2$;
    \item the $r\ell$ points $Q_1\oplus P_i\oplus jT$ ($1\leq i\leq r,\;0\leq j\leq\ell-1$) are pairwise distinct and disjoint from the orbit $\{jT:0\leq j\leq\ell-1\}$.
\end{enumerate}
Define divisors
\[
D = \sum_{i=1}^{r}\sum_{j=0}^{\ell-1} (Q_1\oplus P_i\oplus jT),\qquad \deg(D) = r\ell,
\]
\[
G = s\sum_{j=0}^{\ell-1} jT,\qquad \deg(G) = s\ell.
\]
Then the algebraic geometry code $\mathcal{C}_{\mathscr{L}}(D,G)$ is an $[n,k,n-k+1]$ MDS code with $n=r\ell = 2s\ell$, $k=s\ell$, which is $\ell$-quasi‑cyclic and iso‑dual. In particular, $\mathcal{C}_{\mathscr{L}}(D,G)^{\perp} = \mathbf{v}\mathcal{C}_{\mathscr{L}}(D,G)$ for some vector $\mathbf{v}\in(\mathbb{F}_q^{\ast})^n$.
\end{theorem}
\begin{proof}
The properties of quasi-cyclicity and isoduality are proved analogously to Theorem \ref{th:qc-isodual-translation}.

By Lemma~\ref{lemma:2.5}, the code $\mathcal{C}_{\mathscr{L}}(D,G)$ is MDS if and only if for every choice of $k$ distinct points $P'_1,\dots,P'_k\in \operatorname{supp}(D)$, we have
\[
P'_1\oplus\ldots\oplus P'_k \ne  P_\infty.
\]

Let $P_1',\dots,P_k'$ be any $k=s\ell$ distinct points from $\operatorname{supp}(D)$.  
Each point in $\operatorname{supp}(D)$ is of the form
\[
Q_1\oplus P_i\oplus jT
\]
for some $1\leq i\leq r$ and $0\leq j\leq \ell-1$.

For each $i=1,\dots,r$, let $k_i$ denote the number of selected points of the form $Q_1\oplus P_i\oplus jT$ (with any $j$). Then
\[
0\leq k_i\leq \ell,\qquad \sum_{i=1}^r k_i = k = s\ell.
\]

Summing the selected points gives
\[
P'_1\oplus\ldots\oplus P'_k
=
\Bigl(\sum_{i=1}^r k_i\Bigr) Q_1
\;\oplus\;
\bigoplus_{i=1}^r [k_i]P_i
\;\oplus\;
\lambda T,
\]
where $\lambda$ is the sum of the integers $j$ corresponding to the selected points. Since $s$ and $\ell$ are odd, $\sum k_i = s\ell$ is odd, and because $Q_1$ has order $2$, we have
\[
\Bigl(\sum_{i=1}^r k_i\Bigr) Q_1 = Q_1.
\]

Thus
\[
P'_1\oplus\ldots\oplus P'_k = Q_1 \bigoplus_{i=1}^r [k_i]P_i \oplus \lambda T.
\]

Now observe that each $P_i$ has order coprime to $\ell$ and to $2$, and $T$ has odd order $\ell$. Hence every element of the subgroup generated by the $P_i$ and $T$ has odd order. In particular, $X$ has odd order. Since $Q_1$ has order $2$, it follows that $X \neq Q_1$ (otherwise $Q_1$ would have odd order). Consequently,
\[
P'_1\oplus\ldots\oplus P'_k = Q_1 \bigoplus_{i=1}^r [k_i]P_i \oplus \lambda T \neq  P_\infty.
\]

Thus for every choice of $k$ distinct points in $\operatorname{supp}(D)$, their sum is not $ P_\infty$. By Lemma~\ref{lemma:2.5}, $\mathcal{C}_{\mathscr{L}}(D,G)$ is an MDS code. The Riemann–Roch theorem for elliptic curves gives $\dim \mathscr{L}(G)=\deg(G) = s\ell$, so the parameters are $[r\ell,\;s\ell,\;r\ell-s\ell+1] = [2s\ell,\;s\ell,\;s\ell+1]$.
\end{proof}

\section{Conclusion}
In this paper, we present new results that enable the efficient construction of bases of Riemann–Roch spaces for arbitrary divisors on an elliptic curve, including divisors that are not effective. We have presented an application of this result to the construction of several families of elliptic codes, including: quasi‑cyclic codes, Goppa‑like codes, and MDS codes. As noted in Sections \ref{sec:QC} and \ref{sec:Goppa-like}, subfield subcodes of elliptic codes associated with multipoint effective divisors may have a more random structure compared to the one‑point case $G = kP_\infty$. We note that our result allows for the efficient construction of quasi‑cyclic codes associated with multipoint divisors obtained via the action of the translation group. This significantly increases the possible order of quasi‑cyclicity, which in the classical case $G = kP_\infty$ cannot exceed $6$.

In light of recent results on the construction of distinguishers and structural attacks on Goppa codes \cite{FOPT13,MT23,Mora24,Rand25,Lem25} for certain parameters, it is relevant to consider the possibility of adapting the approaches presented in these works to the cryptanalysis of schemes based on subfield subcodes of elliptic codes.

\bibliography{biblio}
	
\end{document}